\documentclass[10pt,a4paper]{article}

\usepackage{lmodern}
\usepackage[T1]{fontenc}
\usepackage[latin9]{inputenc}
\usepackage{float}
\usepackage{amsmath}
\usepackage{graphicx}
\usepackage{amssymb}
\usepackage{bbm}
\usepackage{esint}
\usepackage{epstopdf}
\usepackage{color}
\usepackage{algorithmic}
\usepackage{algorithm}
\usepackage{abbreviations}
\usepackage{lindsten}
\makeatletter
\floatstyle{ruled}
\newfloat{algorithm}{tbp}{loa}
\floatname{algorithm}{Algorithm}
\usepackage{amsfonts}
\newcommand\Qem{\mathcal{Q}} % EM auxiliary quantity
\newcommand\I[1]{\mathbbm{1}(#1)}
\renewcommand\T{n} % Final time
\newcommand\x{\xi} % Joint state
\renewcommand\S[2]{\mathcal{S}^{(3)}_{#1,#2}}
\renewcommand\P[2]{\mathcal{S}^{(1)}_{#1,#2}}
\newcommand\p[2]{\mathcal{S}^{(2)}_{#1,#2}}
\newcommand\mpr{\alpha} % Mode probabilities

\graphicspath{{./}{../figures/}}

\makeatother
\providecommand{\keywords}[1]{\textbf{\textit{Keywords---}} #1}

\begin{document}
\title{Recursive maximum likelihood identification of\\ jump Markov nonlinear systems}
\author{Emre~{\"O}zkan%
\thanks{Emre {\"O}zkan, Fredrik Lindsten and Fredrik Gustafsson are with the
Division of Automatic Control, Link{\"o}ping University, Link{\"o}ping,
%Department of Electrical Engineering, Link{\"o}ping University, Link{\"o}ping,
Sweden,\texttt{\{emre,lindsten,fredrik\}@isy.liu.se}.},
Fredrik~Lindsten\footnotemark[1],
Carsten Fritsche\thanks{Carsten Fritsche is with IFEN GmbH, Alte Gruber Str. 6, 85586 Poing, Germany, \texttt{carsten@isy.liu.se}}, \\ and
Fredrik~Gustafsson\footnotemark[1]}% <-this % stops a space

\date{October 14, 2013}

\maketitle

\begin{abstract}
In this contribution, we present an online method for joint state and parameter estimation in jump Markov non-linear systems (\jmnls).
State inference is enabled via the use of particle filters which makes the method applicable to a wide range of non-linear models.
To exploit the inherent structure of \jmnls, we design a Rao-Blackwellized particle filter (\rbpf) where the discrete mode is marginalized out analytically.
This results in an efficient implementation of the algorithm and reduces the estimation error variance.
The proposed \rbpf is then used to compute, recursively in time, smoothed estimates of complete data sufficient statistics.
Together with the online expectation maximization algorithm, this enables recursive identification of unknown model parameters.
The performance of the method is illustrated in simulations
and on a localization problem in wireless networks using real data. \\\\
\end{abstract}

% Note that keywords are not normally used for peerreview papers.

% For peer review papers, you can put extra information on the cover
% page as needed:
% \ifCLASSOPTIONpeerreview
% \begin{center} \bfseries EDICS Category: 3-BBND \end{center}
% \fi
%
% For peerreview papers, this IEEEtran command inserts a page break and
% creates the second title. It will be ignored for other modes.

\section{Introduction}
Jump Markov processes have been extensively used in control theory, signal processing,
telecommunications and econometrics for modeling multi-modal behavior of systems (see \eg \cite{Doucet2001a,Logothetis99}
for a brief review of applications). Most studies have focused on a special class of these models, jump Markov linear systems (\jmls),
also known as conditionally linear Gaussian models. In these models, a finite-state Markov chain switches between different linear modes.
The true posterior of a \jmls is a mixture of Gaussians with an exponentially increasing number of components, which is intractable to compute in any realistic scenario.
However, many approximate state inference algorithms have been proposed for \jmls, most of which rely on Kalman filters for computing the conditional
estimates for the linear Gaussian modes. These include the generalized pseudo Bayesian ($\GPB$) approach \cite{Tugnait1982}, the interacting multiple model ($\IMM$) filter \cite{blom1988,Mazor1998} and the Rao-Blackwellized particle filter (\rbpf) \cite{ChenL:2000,DoucetGA:2000}.

In cases when the underlying model has unknown parameters the problem becomes even more challenging,
owing to the coupling between the unknown model parameters and the latent states. There are a number of studies which focuses on the identification and/or adaptation of the \jmls (see \eg \cite{shumway1991dynamic,cinquemani2007general,munir1995adaptive}).
However, these algorithms are not applicable when the dynamic modes of the system are nonlinear.
Such jump Markov \emph{nonlinear} systems (\jmnls) arise in various applications including target
tracking \cite{vo2006,li2003survey}, localization \cite{Nicoli2008,Mihaylova2007},
 econometrics \cite{CarvalhoL:2007}, and audio signal processing \cite{Andrieu03JM}.
Identification of \jmnls is a challenging problem. Indeed, addressing the state inference problem alone is problematic as the various
approximate algorithms mentioned above cannot be used in this setting.
Specially tailored sequential Monte Carlo (\smc) samplers, \ie particle filters, have been proposed
in the literature during the last decade (see the discussion and references below).
These methods can be used for state inference in \jmnls.
However, there has not been much progress made on addressing the joint state and parameter estimation problem for \jmnls.

In this paper, we consider the problem of recursive (\ie online) maximum likelihood identification of \jmnls.
The method that we propose is based on an online expectation maximization (EM) algorithm.
The (batch) EM algorithm \cite{Dempster1977} is one of the most popular methods for maximum likelihood identification of latent
variable models.
It has been applied to a wide range of practical problems in different fields such as statistics, biology and signal processing (see \cite{mclachlan2007} for many examples).

Recent contributions have focused on using EM in an online setting, \ie, when the observations are processed only once and never stored.
The online EM algorithm was initially proposed for hidden Markov models (HMMs) with a finite number of states and observations \cite{Mongillo2008}.
This idea has then been extended to generalized HMMs with, possibly, continuous observations \cite{Cappe2011}.
In \cite{Cappe2009} a particle-filter-based online EM algorithm is proposed for joint state and parameter estimation in general
(possibly non-linear/non-Gaussian) state-space models. This algorithm is further developed in \cite{DelMoral10},
by making use of forward-only smoothing techniques. In \cite{LeCorff2011}, online EM is used to solve the simultaneous localization and mapping problem.
The particle-filter-based online EM \cite{Cappe2009} algorithm is used in \cite{ozkan2012} for estimating the measurement noise distribution in a general state-space models.
The same approach is used specifically for \jmnls in \cite{carsten2012}, without making use of Rao-Blackwellization for the discrete mode variables.

While the algorithms by \cite{Cappe2009,DelMoral10} can be used also for \jmnls, they do not exploit the inherent structure of these models.
As we shall see, this can result in poor performance. Any standard particle filter (see \eg \cite{DoucetJ:2011,Gustafsson:2010a}) can be used for state inference in \jmnls.
However, this can lead to problems due to severe particle degeneracy around mode changes \cite{driessen2005}.
Different improvement strategies have been proposed to address this issue, enabling
efficient use of \smc for \jmnls. In \cite{driessen2005}, particle depletion is prevented by splitting the filtering recursions for the discrete mode and the continuous state, resulting in the IMM particle filter. In \cite{WhiteleyJ:2013,Andrieu03JM}, auxiliary particle filters are used to construct an efficient
sampling strategy. In \cite{Caron2007}, a Markovian prior is assumed for the discrete modes which allows the transition probabilities
to evolve over time, resulting in more robust estimators.

In this paper we propose an alternative modification, namely to make use of a \rbpf.
As mentioned above, the \rbpf is most well known as an algorithm for state inference in conditionally linear Gaussian models,
where one state component is marginalized by running conditional Kalman filters \cite{ChenL:2000,DoucetGA:2000,SchonGN:2005}.
A general \jmnls is not conditionally linear Gaussian, so this approach is not directly applicable. However,
we may still exploit the idea of Rao-Blackwellization, by marginalizing the mode variable using conditional \hmm filters.
This improves the performance over a standard particle filter as the asymptotic variance is reduced \cite{LindstenSO:2011,Chopin:2004}.
Furthermore, by not using particles to represent the mode variable, we are less affected by the degeneracy problems around mode changes as
reported in~\cite{driessen2005}.

To the best of the authors' knowledge, the proposed \rbpf is a novel approach to state inference in \jmnls.
 However, the main contribution of this paper lies in
the adaption of this \rbpf to address the forward-only smoothing problem which lies at the core of online EM.
This further extends the online EM algorithms by \cite{Cappe2009,DelMoral10} to general \jmnls.
The resulting method can be used to estimate the unknown transition probabilities as well as the
unknown model parameters jointly with the state in an online fashion.

%%%%%%%%%%%%%%%%%%%%%%%%%%%%%%%%%%%%%%%%%%%%%%%%%%%%%%%%%%%%%%%%%%%%%%%%%%%%%%%%%%%%%%%%

\section{Expectation Maximization}
The EM algorithm \cite{Dempster1977} is an iterative method which is useful for computing ML estimates, $\widehat{\theta}^{\textrm{ML}}$, of unknown parameters $\theta$ in probabilistic models involving latent variables. Consider the (batch) ML problem,
\begin{equation}
\label{eq:ML_formulation}
\widehat{\theta}^{\textrm{ML}}=\arg\max_{\theta \in \Theta}\,\log p(y_{1:n} ; \theta),
\end{equation}
where $y_{1:n}$ is a collection of $n$ observations and $\Theta$ is the feasible set of parameters.
The idea of the EM algorithm is to separate the original ML estimation problem into two linked problems, denoted by
the expectation (E) step and the maximization (M) step, each of which is hopefully easier to solve than the original problem.
Let $z_{1:n}$ denote the latent variables of the models (for a state-space model, these are typically given by the unobserved state variables).
We then introduce the auxiliary quantity,
\begin{align}
  \nonumber
  \Qem(\theta, {}&\theta^\prime) = \E_{\theta^\prime} \left[ \log p(y_{1:n}, z_{1:n} ; \theta) \mid y_{1:n} \right] \\
  \label{eq:EM:Qdef}
  &= \int \log p(y_{1:n}, z_{1:n} ; \theta) \log p(z_{1:n} \mid y_{1:n} ; \theta^\prime) \,dz_{1:n}.
\end{align}
The auxiliary quantity can be thought of as a proxy for the log-likelihood function. The
EM algorithm is useful when maximization of
$\theta \mapsto \Qem(\theta, \theta^\prime)$, for fixed $\theta^\prime$, is simpler than direct maximization of the log-likelihood,
$\theta \mapsto \log p(y_{1:n} ; \theta)$.
The procedure is initialized at some $\theta^0 \in \Theta$ and then iterates between,
\begin{itemize}
\item \underline{E-Step:} Compute $\Qem(\theta,\theta^{m-1})$.
\item \underline{M-Step:} Compute $\theta^m = \argmax_{\theta \in \Theta} \Qem(\theta,\theta^{m-1})$.
\end{itemize}
At each iteration of the EM algorithm, the parameters are updated so that the value of the log-likelihood is non-decreasing.
The EM algorithm is thus a monotone optimization algorithm.
Furthermore, the resulting sequence $\{\theta^m\}_{m \geq 0}$ will, under weak assumptions, converge
to a stationary point of the likelihood $p(y_{1:n} ; \theta)$ \cite{Wu:1983}.

Note that the auxiliary quantity \eqref{eq:EM:Qdef} is given by the
smoothed estimate of the so called complete data log-likelihood $\log p(y_{1:n}, z_{1:n} ; \theta)$.
This poses an apparent difficulty in using the EM algorithm for solving the online identification problem,
as smoothing is typically an off\-line procedure. However, it has been recognized that this is indeed possible.
The key enabler of the \emph{online EM} algorithm \cite{Mongillo2008,Cappe2011,Cappe2009,DelMoral10} is to make use of forward-only smoothing techniques.
This enables the computation of a stochastic approximation of the auxiliary quantity in an online fashion.
This approximation can then be subsequently used to update the parameters in the M-step at each iteration.
We will discuss how this is done specifically in the context of \jmnls in the subsequent sections.

%%%%%%%%%%%%%%%%%%%%%%%%%%%%%
%%%%%    NEW SECTION   %%%%%%
%%%%%%%%%%%%%%%%%%%%%%%%%%%%%
\section{Jump Markov Nonlinear Systems}
We will derive an online EM algorithm for jump Markov nonlinear systems (\jmnls) in the form,
\begin{subequations}
  \label{eq:jmnls_model}
  \begin{align}
    \label{eq:jmnls_model_a}
    r_t &\sim \Pi(r_t \mid r_{t-1}), \\
    x_{t} &\sim f_{r_t}(x_{t} \mid x_{t-1} ; \theta_{r_t}), \\
    y_t &\sim g_{r_t} (y_{t} \mid x_t ; \theta_{r_t}).
  \end{align}
\end{subequations}
This is a hybrid system with a continuous state variable $x_{t} \in \setX$
and a discrete mode variable $r_t \in \crange{1}{K}$, where $K$ is the number of modes.
The system states $r_t$ and $x_t$ are latent, but observed indirectly through the
measurements $y_t$, taking values in some set $\setY$.
The mode variable follows a
(finite state-space) hidden Markov model (HMM) with transition probabilities
\begin{align}
  \pi_{k\ell} = \Pi(\ell \mid  k) = \probab(r_t = \ell \mid r_{t-1} = k).
\end{align}
The system thus switches between different nonlinear dynamical modes. While in mode $k$,
the transition density function for the state $x_t$ and the likelihood of the measurement $y_t$
are given by $f_k(x_t \mid x_{t-1} ; \theta_k)$ and ${g_k(y_t \mid x_{t} ; \theta_k)}$, respectively.
Each mode $k$ is parameterized by its own set of parameters $\theta_k$. Furthermore,
the transition probabilities $\pi_{k\ell}$ for the mode sequence $\{ r_t \}$ are assumed to be unknown
parameters. By abuse of notation we let $\Pi$ refer to both the transition kernel for $r_t$, as in \eqref{eq:jmnls_model_a},
and the $K\times K$ transition probability matrix with entries $[\Pi]_{k\ell} = \pi_{k\ell}$.
The unknown parameters of the model are thus given by
\begin{align}
  \theta = \left(\{\theta_k\}_{k=1}^K, \Pi \right).
\end{align}
For notational convenience, we assume that the initial state of the system $(x_0, r_0)$ is known.
The generalization to an unknown initial state, exogenous inputs and/or time-inhomogeneous
dynamics is straightforward.
\section{EM Algorithm for JMNLS}
For \jmnls, direct optimization of \eqref{eq:ML_formulation} is typically not possible due to the intractability of computing the likelihood $p(y_{1:n} ; \theta)$.
To address this difficulty, we make use of the EM algorithm. We start the derivation of the online EM algorithm by considering batch EM for \jmnls,
and then continue with the online formulation.

\subsection{Complete data sufficient statistics}
Let the latent variables  $z_{1:n}$ be given by the system states, \ie, $(x_{1:n}, r_{1:n})$. If follows that the complete data likelihood can be factorized as,
\begin{equation}
p(x_{1:n},r_{1:n},y_{1:n};\theta)=\prod_{t=1}^{n}p(x_{t},r_{t},y_{t}|x_{t-1},r_{t-1};\theta)
\label{eq:complete_data_likelihood}
\end{equation}
In the following, we focus on the complete data sufficient statistics formulation of the EM algorithm \cite{Cappe2009,mclachlan2007}.
It is assumed that the nonlinear dynamical system corresponding to each mode belongs to the curved exponential family.
That is, for each $k\in\range{1}{K}$ we have
\begin{align}
  \nonumber
  &g_k(y_t \mid x_{t} ; \theta_k) f_k(x_t \mid x_{t-1} ; \theta_k) \\
  &= C_k \exp\left( \left<\psi_k(\theta_k), s_{k,t}(y_t, x_t, x_{t-1}) \right> - A_k(\theta_k) \right),
\end{align}
where $C_k$ may depend on $y_t$, $x_t$ and $x_{t-1}$, but is independent of $\theta_k$;
$\left<\cdot, \cdot\right>$ denotes inner product; $\psi_k(\theta_k)$ is the natural parameter;
$s_{k,t}(y_t,x_{t},x_{t-1})$ is the complete data sufficient statistic and $A_k(\theta_k)$ denotes the log-partition function.

Furthermore, we assume that there exist unique maximizers of the complete data likelihoods.
That is, for each $k = \range{1}{K}$, there exists a mapping $\Lambda_k(\mathcal{S}) \mapsto \theta_k$ given by,
\begin{align}
  \Lambda_k(\mathcal{S}) = \argmax_{\theta_k \in \Theta_k} \{ \left<\psi_k(\theta_k), \mathcal{S}) \right> - A_k(\theta_k)   \},
\end{align}
where $\Theta_k$ is the feasible set for the parameter $\theta_k$.
In Appendix~\ref{app:noises}, we provide the explicit expressions for these mappings for the special case of jump Markov Gaussian
systems with unknown noise parameters.

%%%%%%%%%%%%%%%

In order to compute the auxiliary quantity \eqref{eq:EM:Qdef}, we make use of the indicator function $\I{\cdot}$ and
write the logarithm of the complete data likelihood \eqref{eq:complete_data_likelihood} as (omitting constant terms),
\begin{align}
  \nonumber
  &\log p(x_{1:\T}, r_{1:\T}, y_{1:\T} ; \theta) = \sum_{t=1}^\T \log \Pi(r_t \mid r_{t-1}) \\
  \nonumber
  &+ \sum_{t=1}^\T \log \left( g_{r_t}(y_t \mid x_{t} ; \theta_{r_t}) f_{r_t}(x_t \mid x_{t-1} ; \theta_{r_t}) \right) \\
  \nonumber
  &= \sum_{k=1}^K \sum_{\ell = 1}^K \sum_{t=1}^\T  \log \left( \pi_{k\ell} \right) \I{r_t = \ell, r_{t-1} = k} \\
  &+ \sum_{k=1}^K \sum_{t=1}^\T \I{r_t = k} \left( \left<\psi_k(\theta_k), s_{k,t}(y_t, x_t, x_{t-1}) \right> - A_k(\theta_k) \right). %+ \text{const.}
\end{align}
The auxiliary quantity of the EM algorithm can be written as,
\begin{align}
  \nonumber
  \Qem(\theta, \theta^\prime) &=
  \E_{\theta^\prime}[\log p(x_{1:\T}, r_{1:\T}, y_{1:\T} ; \theta) \mid y_{1:\T}] \\   \label{eq:EM_JMNLS:Qfunc_expr}
  &= \sum_{k=1}^K \sum_{\ell = 1}^K \P{k\ell}{\T} \log \pi_{k\ell}
+ \sum_{k=1}^K \left( \left<\psi_k(\theta_k), \S{k}{\T} \right> - A_k(\theta_k) \p{k}{n} \right),
\end{align}
where we have introduced the sufficient statistics,
\begin{subequations}
  \label{eq:onlinem_smoothedstatistics}
  \begin{align}
    \P{k\ell}{\T} &= \sum_{t=1}^\T \E_{\theta^\prime}[\I{r_t = \ell, r_{t-1} = k} \mid y_{1:\T}], \\
    \p{k}{\T} &= \sum_{t=1}^\T \E_{\theta^\prime}[\I{r_t = k} \mid y_{1:\T}], \\
    \S{k}{\T} &= \sum_{t=1}^\T \E_{\theta^\prime}[\I{r_t = k} s_{k,t}(y_t, x_t, x_{t-1}) \mid y_{1:\T}],
  \end{align}
for $k,\ell = \range{1}{K}$.
\end{subequations}

%%%%%%%%%%%%%%%

For the M-step, we need to maximize \eqref{eq:EM_JMNLS:Qfunc_expr} \wrt $\theta$. However, we note
that the objective function is separable across the modes. That is, we can maximize each term of the second sum independently \wrt $\theta_k$.
Furthermore, the transition probability matrix $\Pi$ only enters the first sum in \eqref{eq:EM_JMNLS:Qfunc_expr}.
This term can thus be maximized, under the constraints $\pi_{k\ell} \geq 0$ and $\sum_{\ell = 1}^K \pi_{k\ell} = 1$, by using
standard formulae for HMMs (see \eg \cite{CappeMR:2005}).
The M-step can be computed as follows:
\begin{subequations}
  \label{eq:EM_JMNLS:parameterupdate}
  \begin{align}
    \widehat \theta_k &= \Lambda_k(\S{k}{\T} / \p{k}{\T}), & k&=\range{1}{K}, \\
    \label{eq:EM_JMNLS:parameterupdate_b}
    \widehat \pi_{k\ell} &= \frac{\P{k\ell}{\T} }{ \sum_{j=1}^K \P{kj}{\T} } , & k, \ell&=\range{1}{K}.
  \end{align}
\end{subequations}
Note that it is possible to extend the model to account for common but unknown parameters across different modes.
Furthermore, constraints on the parameters in the original ML formulation carry over to the M-step of the EM algorithm. This makes the algorithm suitable for constrained parameter estimation problems.

%%%%%%%%%%%%%%%%%%%%%%%%%%%%%
%%%%%    NEW SECTION   %%%%%%
%%%%%%%%%%%%%%%%%%%%%%%%%%%%%
\subsection{Online-EM for \jmnls}
A closer look at \eqref{eq:onlinem_smoothedstatistics} reveals that the EM algorithm requires the computation of smoothed additive functionals.
In an offline implementation, standard forward/backward or two-filter smoothers may be used to compute these
smoothed estimates; see \eg \cite{LindstenS:2013} for a recent survey.
However, for online EM, the smoothed functionals need to be computed online. For the case of
additive functionals, this is in fact possible by using so called forward-only smoothing techniques (see \eg \cite{Cappe2005,DelMoral10})
which are based on dynamic programming.

For notational simplicity, we use the joint state variable $\x_t = (r_t, x_t)$.
Let,
\begin{subequations}
  \begin{align}
    s_t^{(1)}(\x_t, \x_{t-1}) &= \vectorize( \{\I{r_t = \ell, r_{t-1} = k}\}_{k,\ell = 1}^K), \\
    s_t^{(2)}(\x_t, \x_{t-1}) &= \vectorize(\{ \I{r_t = k} \}_{k = 1}^K), \\
    s_t^{(3)}(\x_t, \x_{t-1}) &= \vectorize(\{ \I{r_t = k} s_{k,t}(y_t, x_t, x_{t-1})\}_{k = 1}^K),
  \end{align}
\end{subequations}
where $\vectorize(\cdot)$ is the vectorization operator, which stacks the elements of a set in a vector
(using some convenient ordering) and where we have removed the dependence on $y_t$ in the notation for brevity. Furthermore, let
\begin{align}
  s_t(\x_t, \x_{t-1}) =
  \begin{pmatrix}
    s_t^{(1)}(\x_t, \x_{t-1}) \\ s_t^{(2)}(\x_t, \x_{t-1}) \\ s_t^{(3)}(\x_t, \x_{t-1})
  \end{pmatrix}.
\end{align}
It follows that \eqref{eq:onlinem_smoothedstatistics} can be written compactly as,
\begin{align}
  \label{eq:fsm_smoothedS}
  \mathcal{S}_\T = \E_{\theta^\prime}[S_\T(\x_{0:\T}) \mid y_{1:\T}].
\end{align}
with
\begin{align}
  \label{eq:fsm_Sdef}
  S_\T(\x_{0:\T}) = \sum_{t=1}^\T s_t(\x_{t-1}, \x_t).
\end{align}
Consider the intermediate quantity $T_t(\x_t) \triangleq \E_{\theta^\prime}[ S_t(\x_{0:t}) \mid \x_t, y_{0:t}]$.
Note that $T_t(\x_t)$ is a function of the joint state $\x_t$. From the tower property of conditional expectation, it follows that
\begin{align}
  \label{eq:fsm_smoothedSfromT}
  \mathcal{S}_t = \E_{\theta^\prime}[T_t(\x_t) \mid y_{1:t}] = \int T_t(\x_t) p(\x_t \mid y_{1:t} ; \theta^\prime) \, d\x_t.
\end{align}
That is, the smoothed additive functional \eqref{eq:fsm_smoothedS} is given by the
filtered estimate of $T_t(\x_t)$. Furthermore, the additive form \eqref{eq:fsm_Sdef} allows
us to express $T_t$ recursively,
\begin{align}
  \nonumber
  T_t(\x_t) &= \int \left[ T_{t-1}(\x_{t-1}) + s_t(\x_{t-1}, \x_t) \right] \\
  \label{eq:fsm_recursiveT}
  &\hspace{4em}\times p(\x_{t-1} \mid \x_t, y_{1:t-1}; \theta^\prime) \, d\x_{t-1},
\end{align}
with $T_0(\x_t)\equiv 0$.

The online EM algorithm exploits the recursive form in \eqref{eq:fsm_recursiveT}. At each time step $t$, the intermediate
quantity $T_t(\x_t)$ is updated and a new parameter estimate $\widehat\theta^t$ is computed according to \eqref{eq:EM_JMNLS:parameterupdate}.
Since the parameters of the model are updated on the fly, a stochastic approximation type of forgetting is
used to update the intermediate quantity. That is, we update $T_t(\x_t)$ at each iteration according to,
\begin{align}
  \nonumber
  T_t(\x_t) \gets \int &\left[(1-\gamma_t) T_{t-1}(\x_{t-1}) + \gamma_t s_t(\x_{t-1}, \x_t) \right] \\
  \label{eq:onlineEMrecursiveT}\
  &\hspace{2em}\times p(\x_{t-1} \mid \x_t, y_{1:t-1}; \widehat\theta^{t-1}) \, d\x_{t-1},
\end{align}
where $\widehat\theta^{t-1}$ is the current parameter estimate and $\{\gamma_{t}\}_{t\ge1}$ is a sequence of decreasing
step-sizes, satisfying the stochastic approximation requirements
$\sum_{t\ge1}\gamma_t=\infty$ and $\sum_{t\ge1}\gamma_t^2<\infty$.
See \cite{Cappe2009,DelMoral10,Cappe2011} for further discussion on the online EM algorithm.

%%%%%%%%%%%%%%%%%%%%%%%%%%%%%
%%%%%    NEW SECTION   %%%%%%
%%%%%%%%%%%%%%%%%%%%%%%%%%%%%

\section{\smc Implementation}
Exact computation of the smoothed statistics in \eqref{eq:fsm_smoothedSfromT} and \eqref{eq:onlineEMrecursiveT} is
not possible in general for a \jmnls. We now turn to computational methods based on \smc to approximate
these quantities.

\subsection{Rao-Blackwellized particle filter for \jmnls}
Rao-Blackwellization (or marginalization) is a key step in efficient implementation of particle filters. Much previous work has been focused on
conditionally linear Gaussian models, where one state component is marginalized by running conditional Kalman filters \cite{ChenL:2000,DoucetGA:2000,SchonGN:2005}.
This is not possible for the case of \jmnls, as the dynamical modes are themselves nonlinear. Instead, we propose to utilize Rao-Blackwellization by marginalizing the discrete state variable using conditional \hmm filters.

We start by considering the filtering problem, \ie, to compute the
filtering densities $p(x_t, r_t \mid y_{1:t})$ for $t = \range{1}{\T}$.
For brevity, we drop the unknown parameter $\theta$ from the notation throughout this section.
To be able to marginalize the mode variable $r_t$, we consider the extended target density,
\begin{align}
  \label{eq:rbpf_factorization}
  p(x_{1:t}, r_t \mid y_{1:t}) = p(r_t \mid x_{1:t}, y_{1:t}) p(x_{1:t} \mid y_{1:t}).
\end{align}
Note that the filtering density is given as a marginal of the above \pdf. The second factor
is approximated using a PF, which is represented by a set of $\Np$ weighted particles $\{x_{1:t}^i, w_t^i\}_{i=1}^\Np$,
each being a state trajectory $x_{1:t} \in \setX^t$. The particles define a point-mass approximation in the form,
\begin{align}
  \label{eq:rbpf_empdist_x}
  \widehat p^\Np(x_{1:t} \mid y_{1:t}) = \sum_{i=1}^\Np w_t^i \delta_{x_{1:t}^i} (x_{1:t}),
\end{align}
where $\delta_z(x)$ is a Dirac point-mass located at the point $z$.
Conditionally on $x_{1:t}$, the mode variable $r_{t}$ follows a finite state-space \hmm. Hence,
the conditional density of $r_t$ in \eqref{eq:rbpf_factorization} is available by running a conditional \hmm filter.
This allows us to compute the mode probabilities,
\begin{align}
  \label{eq:rbpf_empdist_r}
  \mpr_{t\mid t}^i(\ell) &\triangleq \probab(r_t = \ell \mid x_{1:t}^i, y_{1:t}),
\end{align}
for $\ell = \range{1}{K}$ and $i = \range{1}{\Np}$.

At time $t = 0$ we have $\mpr_{0\mid 0}^i(\ell) \equiv \I{r_0 = \ell}$, since the initial state is assumed known
(as before, the generalization to an unknown initial state is straightforward).
Additionally, we set $x_0^i = x_0$ and $w_0^i = 1/\Np$ for $i = \range{1}{\Np}$.
Assume that we have obtained
approximations according to \eqref{eq:rbpf_empdist_x} and \eqref{eq:rbpf_empdist_r}
for time $t-1$, represented by the particle system
\begin{align}
  \label{eq:rbpf_wps}  \{x_{1:t-1}^i, w_{t-1}^i, \mpr_{t-1\mid t-1}^i(\cdot) \}_{i=1}^\Np.
\end{align}
Here, we have included the conditional filtering probabilities for the mode variable in the particle system.
For notational convenience, $\mpr_{t-1\mid t-1}^i(\cdot)$ refers to the set $\{\mpr_{t-1\mid t-1}^i(\ell)  \}_{\ell=1}^K$
(and similarly for the prediction probabilities).
We will now derive the update equations for the \rbpf,
and see how to propagate this particle system to time $t$.

First, as for any \smc sampler, resampling is conducted to rejuvenate the particles and reduce the effects
of degeneracy \cite{DoucetJ:2011}. Resampling does not have to be done at every iteration of the algorithm.
Instead, we can choose to resample only when, say, the effective sample size
drops below some user-defined threshold (see \eg \cite{Cappe2005,Liu:2001}). In either case, let
\begin{align}
  \label{eq:rbpf_wps_resampled}
  \{\widetilde x_{1:t-1}^i, \widetilde w_{t-1}^i, \widetilde \mpr_{t-1\mid t-1}^i(\cdot) \}_{i=1}^\Np,
\end{align}
refer to the weighted particle system obtained after the resampling step of the algorithm. Note that \eqref{eq:rbpf_wps_resampled} is
identical to \eqref{eq:rbpf_wps} if no resampling is done at time $t-1$.

Consider now the time update of the conditional \hmm filter. Analogously to \eqref{eq:rbpf_empdist_r}, we define the
predictive mode probabilities (\wrt the resampled particle trajectories),
\begin{align}
  \widetilde\mpr_{t\mid t-1}^i(\ell) &\triangleq \probab(r_{t} = \ell \mid \widetilde x_{1:t-1}^i, y_{1:t-1}),
\end{align}
By using the Markov property of the mode sequence we can write
\begin{align}
  \nonumber
  &\probab(r_{t} = \ell, r_{t-1} = k \mid x_{1:t-1}, y_{1:t-1}) \\
  &= \Pi(r_{t} = \ell \mid r_{t-1} = k) \probab(r_{t-1} = k \mid x_{1:t-1}, y_{1:t-1}).
\end{align}
By marginalizing the above expression over $r_{t-1}$ we thus get,
\begin{align}
  \label{eq:rbpf_alpha_predict}
  \widetilde \mpr_{t\mid t-1}^i(\ell) = \sum_{k = 1}^K \pi_{k\ell} \widetilde \mpr_{t-1 \mid t-1}^{i}(k),
\end{align}
for $\ell = \range{1}{K}$ and $i = \range{1}{\Np}$.

Next, we consider updating the continuous state variable.
To extend the particle trajectories to time $t$, we draw new samples from some proposal kernel according to
\begin{align}
  \label{eq:rbpf_proposal}
  x_t^i \sim q_t(x_t \mid \widetilde x_{1:t-1}^{i}, y_{1:t}),
\end{align}
and set $x_{1:t}^i = (\widetilde x_{1:t-1}^i, x_t^i)$ for $i = \range{1}{\Np}$.
Given the new particles and the current measurement $y_t$,
we can compute the updated mode probabilities \eqref{eq:rbpf_empdist_r}.
This constitutes the measurement update of the conditional \hmm filter.
Note, however, that the continuous state $x_t$ carries information about $r_t$, and thus serves as an ``extra measurement''.
Let us define the quantities,
\begin{align}
  \nonumber
  \gamma_t^i(r_t) &\triangleq p(y_t, x_t^i, r_t \mid \widetilde x_{1:t-1}^i, y_{1:t-1}) \\
  \label{eq:rbpf_gammadef}
  &= g_{r_t} (y_t \mid x_t^i) f_{r_t}(x_t^i \mid \widetilde x_{t-1}^{i}) \widetilde\mpr_{t\mid t-1}^{i}(r_{t}).
\end{align}
Since $p(r_t \mid x_{1:t}, y_{1:t}) \propto p(y_t, x_t, r_t \mid x_{1:t-1}, y_{1:t-1})$,
it follows that,
\begin{align}
  \label{eq:rbpf_alpha_update}
  \mpr_{t\mid t}^i(\ell) = \frac{ \gamma_t^i(\ell) }{ \sum_{k=1}^K \gamma_t^i(k) },
\end{align}
for $\ell = \range{1}{K}$ and $i = \range{1}{\Np}$.

Finally, to account for the discrepancy between the target and the proposal distributions,
the particles are assigned importance weights according to,
\begin{align}
  \label{eq:rbpf_weights}
  w_t^i \propto \frac{p(x_t^i, y_t \mid \widetilde x_{1:t-1}^i, y_{1:t-1})}{q_t(x_t^i \mid \widetilde x_{1:t-1}^i, y_{1:t})} \widetilde w_{t-1}^i.
\end{align}
The numerator of this expression is given by marginalizing \eqref{eq:rbpf_gammadef} over $r_t$, \ie,
by the normalization constant $\sum_{k=1}^K \gamma_t^i(k)$.
As in a standard PF, the weights are then normalized to sum to one.

It is worth to emphasize that standard modifications from the \smc literature may be used
together with the \rbpf, \eg resampling with adjustment weights \cite{PittS:1999} or
incorporating \mcmc moves in the sampler \cite{GilksB:2001}. It can also be of interest, from an implementation point of view,
to note that the ``bootstrap proposal''  for the \rbpf is given by the mixture distribution,
\begin{align}
  p(x_t \mid \widetilde x_{1:t-1}^{i}, y_{1:t-1}) = \sum_{k=1}^K f_{k}(x_t \mid \widetilde x_{t-1}^{i}) \widetilde\mpr_{t\mid t-1}^{i}(k).
\end{align}
We summarize the \rbpf in Algorithm~\ref{alg:rbpf}.

\begin{algorithm}
  \caption{\rbpf for \jmnls (at time $t$)}
  \label{alg:rbpf}
  \begin{enumerate}
  \item \textbf{Input: }   $\{x_{1:t-1}^i, w_{t-1}^i, \mpr_{t-1\mid t-1}^i(\cdot) \}_{i=1}^\Np$.
  \item \textbf{Resampling: } Optionally resample the particles, or retain the previous particle system. Let
    $\{\widetilde x_{1:t-1}^i, \widetilde w_{t-1}^i, \widetilde \mpr_{t-1\mid t-1}^i(\cdot) \}_{i=1}^\Np$
    denote the result.
  \item \textbf{For $i = 1$ to $\Np$ do:}
    \begin{enumerate}
    \item Compute $\{ \widetilde \mpr_{t\mid t-1}^i(\ell) \}_{\ell=1}^K$ according to \eqref{eq:rbpf_alpha_predict}.
    \item Draw $x_t^i \sim q_t(x_t \mid \widetilde x_{1:t-1}^{i}, y_{1:t})$ and set $x_{1:t}^i = (\widetilde x_{1:t-1}^i, x_t^i)$.
    \item Compute $\{ \gamma_t^i(\ell) \}_{\ell=1}^K$ according to \eqref{eq:rbpf_gammadef}.
    \item Compute $\{ \mpr_{t\mid t}^i(\ell) \}_{\ell=1}^K$ according to \eqref{eq:rbpf_alpha_update}.
    \item Set
      \begin{align*}
        w_t^{\prime,i} = \frac{ \sum_{k=1}^K \gamma_t^i(k)  }{q_t(x_t^i \mid \widetilde x_{1:t-1}^i, y_{1:t})} \widetilde w_{t-1}^i.
      \end{align*}
    \end{enumerate}
  \item[] \textbf{End for}
  \item \textbf{Normalize:} $ w_t^{i} = w_t^{\prime,i}/ \sum_{j=1}^\Np w_t^{\prime,j}$, $i = \range{1}{\Np}$.
  \item \textbf{Output: }   $\{x_{1:t}^i, w_{t}^i, \mpr_{t\mid t}^i(\cdot) \}_{i=1}^\Np$.
  \end{enumerate}
\end{algorithm}

%%%%%%%%%%%%%%%%%%%%%%%%%%%%%
%%%%%    NEW SECTION   %%%%%%
%%%%%%%%%%%%%%%%%%%%%%%%%%%%%
%\subsection{Forward smoothing using RBPF}
\subsection{RBPF-based online-EM}
Using \eqref{eq:fsm_smoothedSfromT} and \eqref{eq:onlineEMrecursiveT}, we seek a recursive
approximation of $\mathcal{S}_t$ based on the \rbpf.
For each particle in the system $\{x_t^i, w_t^i\}_{i=1}^N$ we will compute an approximation $\widehat T_t^i(r_t)$ of the
intermediate quantity $T_t(x_t, r_t)$, \ie,
\begin{align}
  \label{eq:fsm:rbpf_approxT}
   \widehat T_t^i(\ell) &\approx T_t(x_t^i,r_t = \ell)
\end{align}
for $i = \range{1}{\Np}$ and $\ell = \range{1}{K}$. Given these quantities, it follows from \eqref{eq:fsm_smoothedSfromT}
that we can approximate $\mathcal{S}_t$ by the \rbpf, according to,
\begin{align}
  \mathcal S_t \approx \widehat{\mathcal{S}}_t^\Np \triangleq \sum_{i = 1}^\Np \sum_{\ell=1}^K w_t^i \mpr_{t\mid t}^i(\ell) \widehat T_t^{i}(\ell).
\end{align}
It remains to compute the intermediate quantities \eqref{eq:fsm:rbpf_approxT}. From the updating equation \eqref{eq:onlineEMrecursiveT},
we note that this requires us to compute an expectation under the so called \emph{backward kernel}
$p(\x_{t-1} \mid \x_t, y_{1:t-1})$. The key step in computing \eqref{eq:fsm:rbpf_approxT} is thus
to find an approximation of the backward kernel based on the \rbpf particles.
We will consider two different approaches, leading to different algorithms.
The first is more accurate, but its computational cost scales quadratically with the number of particles.
The latter leads to a cruder approximation, but its computational cost scales only linearly with the number of particles.

Note that \eqref{eq:onlineEMrecursiveT} can be written as
\begin{align}
  \nonumber
  T_t(\x_t)  &\gets \sum_{r_{t-1}} \int \left[ (1-\gamma_t) T_{t-1}(\x_{t-1}) + \gamma_t s_t(\x_{t-1}, \x_t) \right] \\
  \label{eq:fsm:rbpf_recursiveT}
   &\hspace{2em}\times p(x_{1:t-1}, r_{t-1} \mid \x_t, y_{1:t-1}) \, dx_{1:t-1},
\end{align}
where we have extended the integration to the complete trajectory $x_{1:t-1}$ (which does not alter the value of the integral).
The extended backward kernel density can be written as,
\begin{align}
  \nonumber
  &p(x_{1:t-1}, r_{t-1} \mid x_t, r_t, y_{1:t-1}) \propto f_{r_t}(x_t \mid x_{t-1}) \Pi(r_t \mid r_{t-1}) \\
  \label{eq:fsm_rbpffs_bwdkernel}
  &\hspace{2em}\times p(r_{t-1} \mid x_{1:t-1}, y_{1:t-1}) p(x_{1:t-1} \mid y_{1:t-1}).
\end{align}
By plugging in the \rbpf approximations \eqref{eq:rbpf_empdist_x} and \eqref{eq:rbpf_empdist_r} we get,
\begin{align}
  \nonumber
  p(&x_{1:t-1}, r_{t-1}=k \mid x_t^i, r_t = \ell, y_{1:t-1}) \\
  \label{eq:fsm_rbpffs_bwdkernelapprox1}
  &\approx \sum_{j=1}^\Np \frac{ \bar w_t^{i,j}(k,\ell)}{ \sum_{u=1}^\Np \sum_{m=1}^K \bar w_t^{i,u}(m,\ell) }  \delta_{x_{1:t-1}^j} (x_{1:t-1}),
\end{align}
with
\begin{align}
  \bar w_t^{i,j}(k,\ell) =  f_\ell(x_t^i \mid x_{t-1}^j) \pi_{k\ell} \alpha_{t-1\mid t-1}^j(k) w_{t-1}^j.
\end{align}
By using this approximation of the backward kernel, we obtain the following update equation for the intermediate quantities,
\begin{align}
  \nonumber
  &\widehat T_t^i(\ell) = \sum_{j=1}^\Np \sum_{k=1}^K  \Bigg( \frac{ \bar w_t^{i,j}(k,\ell)}{ \sum_{u=1}^\Np \sum_{m=1}^K \bar w_t^{i,u}(m,\ell) } \\
  \label{eq:rbpffs_TN2}
  &\times \left[ (1-\gamma_t) \widehat T_{t-1}^{j}(k) + \gamma_t s_t(x_{t-1}^j, r_{t-1} = k, x_t^i, r_t = \ell) \right] \Bigg).
\end{align}
The recursion is initialized by $\widehat T_0^i(\ell) \equiv 0$.

The computational complexity of computing \eqref{eq:rbpffs_TN2} for $\ell = \range{1}{K}$ and $i = \range{1}{\Np}$
is $\Ordo(K^2\Np^2)$. One way to reduce the computational complexity of the forward smoother, is to rely on path-based smoothing.
The backward kernel approximation \eqref{eq:fsm_rbpffs_bwdkernelapprox1} can be thought of as considering
all particles at time $t-1$ as possible ancestors to each particle at time $t$. An alternative
is to only consider the ``actual'' ancestors. Recall that $\{ \widetilde x_{t-1} \}_{i=1}^\Np$ are the
resampled particles at time $t-1$ and that $x_t^i$ originates from $\widetilde x_{t-1}^i$ as in \eqref{eq:rbpf_proposal}.
This suggests to approximate \eqref{eq:fsm_rbpffs_bwdkernel} according to,
\begin{align}
  \nonumber
  p(&x_{1:t-1}, r_{t-1} = k \mid x_t^i, r_t = \ell, y_{1:t-1}) \\
  \label{eq:rbpffs_bwdkernelapprox2}
  &\approx \frac{ \pi_{k\ell} \widetilde\alpha_{t-1\mid t-1}^{i}(\ell)   }
  {\sum_{m=1}^K \pi_{m\ell} \widetilde\alpha_{t-1\mid t-1}^{i}(m)}  \delta_{\tilde{x}_{1:t-1}^i} (x_{1:t-1}).
\end{align}
Note that the factors $f_\ell(x_t^i \mid \widetilde x_{t-1}^i)$ cancel when normalizing the distribution.
Let $\{\widetilde T_{t-1}^i(\ell)\}_{i = 1}^\Np$ be the set of resampled intermediate statistics, arising
from resampling $\{\widehat T_{t-1}^i(\ell)\}_{\ell = 1}^\Np$ along with the particles at time $t-1$.
By using \eqref{eq:rbpffs_bwdkernelapprox2} in \eqref{eq:fsm:rbpf_recursiveT} we get the alternative updating equation
for the intermediate quantities,
\begin{align}
  \nonumber
  &\widehat T_t^i(\ell) = \sum_{k=1}^K \Bigg( \frac{ \pi_{k\ell} \widetilde\alpha_{t-1\mid t-1}^{i}(k) }{ \sum_{m=1}^K \pi_{m\ell} \widetilde\alpha_{t-1\mid t-1}^{i}(m) } \\
  \label{eq:rbpffs_TN}
  &\times \left[ (1-\gamma_t) \widetilde T_{t-1}^{i}(k) + \gamma_t s_t(\widetilde x_{t-1}^{i}, r_{t-1} = k, x_t^i, r_t = \ell) \right] \Bigg).
\end{align}
As before, the recursion is initialized with $\widehat T_0^i(\ell) \equiv 0$.
The computational complexity of computing these quantities for $\ell = \range{1}{K}$ and $i = \range{1}{\Np}$ is $\Ordo(K^2\Np)$.
The price we pay for the reduced computational complexity is a cruder approximation of the backward kernel.
Indeed, since \eqref{eq:rbpffs_bwdkernelapprox2} relies on path-based smoothing,
it will suffer from path degeneracy. However, it turns out that the effect of the degeneracy
is not as bad as one might first think, due to the inherent forgetting factor in the online EM algorithm.
Still, as we shall see in Section~\ref{sec:experiments}, \eqref{eq:rbpffs_TN} leads to a larger variance of the resulting parameter estimates
than \eqref{eq:rbpffs_TN2}.

We summarize the RBPF-based online EM algorithm for \jmnls in Algorithm~\ref{alg:rbpf_onlinem}.

\begin{algorithm}
  \caption{Online EM for \jmnls (at time $t$)}
  \label{alg:rbpf_onlinem}
  \begin{enumerate}
  \item \textbf{Filter update:}
    \begin{enumerate}
    \item Parameterize the model with $\widehat\parameter^{t-1}$.
    \item Run one step of the \rbpf (Algorithm~\ref{alg:rbpf}).
    \end{enumerate}
  \item \textbf{Parameter update:}
    \begin{enumerate}
    \item Compute $\{\widehat T_t^i(\cdot) \}_{i=1}^\Np$ according to
      \begin{itemize}
      \item \textit{(Forward smoothing):} \eqref{eq:rbpffs_TN2}
      \item \textit{(Path-based smoothing):} \eqref{eq:rbpffs_TN}
      \end{itemize}
    \item Compute $\widehat{\mathcal{S}}_t^\Np =\sum_{i = 1}^\Np \sum_{\ell=1}^K w_t^i \mpr_{t\mid t}^i(\ell) \widehat T_t^{i}(\ell)$.
    \item Update the parameter $\widehat\theta^t$ according to \eqref{eq:EM_JMNLS:parameterupdate}.
    \end{enumerate}
  \end{enumerate}
\end{algorithm}

\section{Experimental Results}\label{sec:experiments}%
%%%%%%%%%%%%%%%%%%%%%%%%%%%%%%%%%%%%%%%%%%%%%%%%
%%%%%%%% MEASUREMENT NOISE SIMULATION %%%%%%%%%%
%%%%%%%%%%%%%%%%%%%%%%%%%%%%%%%%%%%%%%%%%%%%%%%%
\subsection{Simulations}
In this section we compare the performance of different implementations of the online $\EM$ algorithm
on a benchmark model and illustrate the gain in Rao-Blackwellization.
Consider the following modified benchmark model:
\begin{subequations}
\label{eq:benchmark_model}
\begin{align}
x_{t}&=\frac{1}{2}x_{t-1}+25\,\frac{x_{t-1}}{1+x_{t-1}^{2}}+8\cos(1.2\,t) + v_t,\\
y_{t}&=\frac{x_{t}^2}{20}+e_t^{(r_t)},
\end{align}
\end{subequations}
where $v_t \sim \N(0,1)$ and where the measurement noise is governed by a $2$-state Markov chain $r_t \in \{1,2\}$. The mode-dependent measurement noise
is assumed to be Gaussian distributed, $e_{t}^{(k)} \sim \mathcal{N}(\mu_{\textrm{e},k},\Sigma_{\textrm{e},k})$, $k=1,2$.
The mode-dependent mean and variance as well as the transition probabilities of the Markov chain are assumed unknown, \ie,
the parameters of the model are $\theta = \left(\{\theta_k\}_{k=1}^2, \Pi \right)$.
Here, $\Pi$ refers to the $2\times 2$ transition probability matrix (TPM) with entries $[\Pi]_{k\ell} = \pi_{k\ell}$ and  $\theta_{k}=\{\mu_{\textrm{e},k},\Sigma_{\textrm{e},k}\}$. The model parameter values used in the simulations are summarized in
Table~\ref{table_simulation_parameters}.
\begin{table}[!t]
\renewcommand{\arraystretch}{1.3}
\caption{Benchmark model parameters}
\label{table_simulation_parameters}
\centering
\begin{tabular}{c|c||c|c}
\hline
\bfseries Parameter & \bfseries Value & \bfseries Parameter & \bfseries Value\\
\hline\hline
$\pi_{11}$ & $0.95$ & $\pi_{22}$ & $0.8$ \\
$\mu_{\textrm{e},1}$ & $0$ & $\mu_{\textrm{e},2}$ & $3$\\
$\Sigma_{\textrm{e},1}$ & $1$ & $\Sigma_{\textrm{e},2}$ & $4$\\
\hline
\end{tabular}
\end{table}

\begin{figure*}[ptb]
  \centering
  \includegraphics[width=0.25\linewidth]{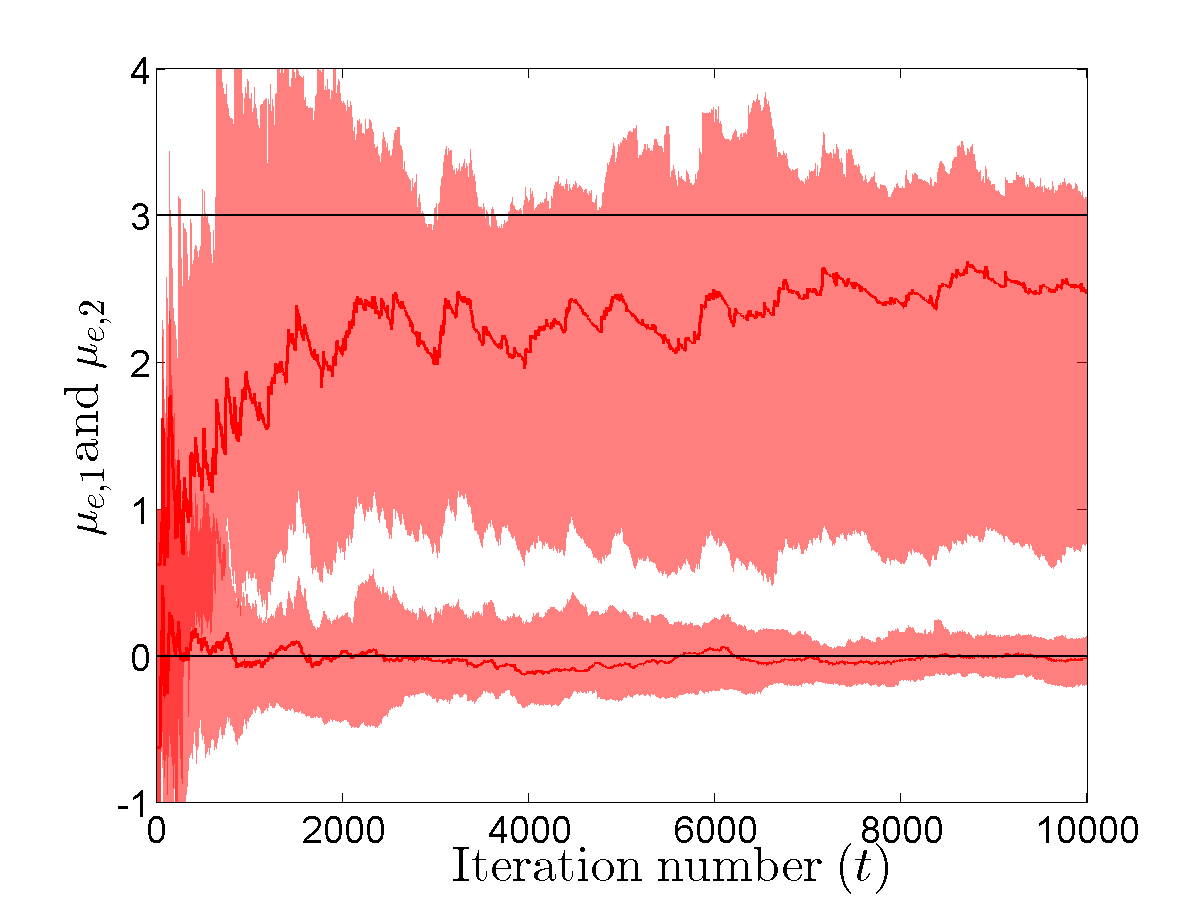}%
  \includegraphics[width=0.25\linewidth]{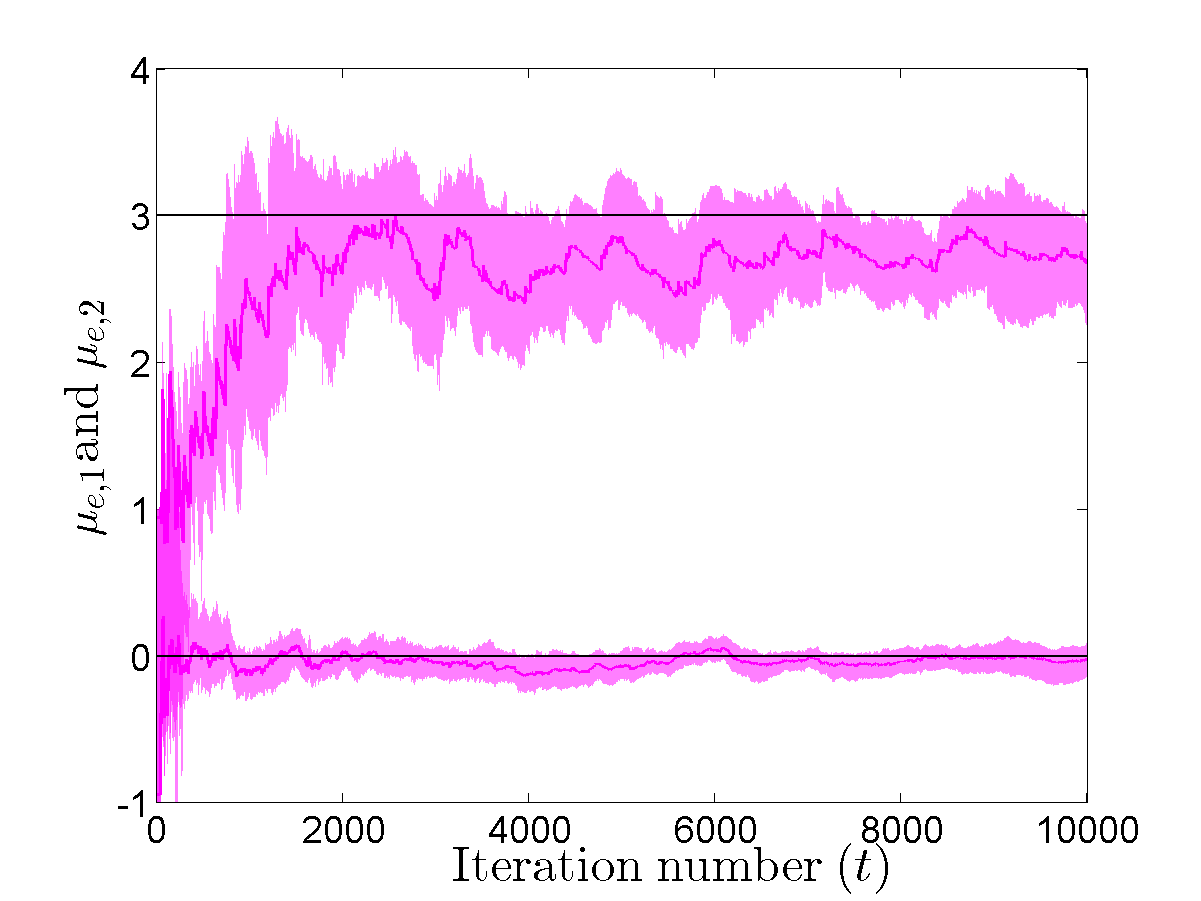}%
  \includegraphics[width=0.25\linewidth]{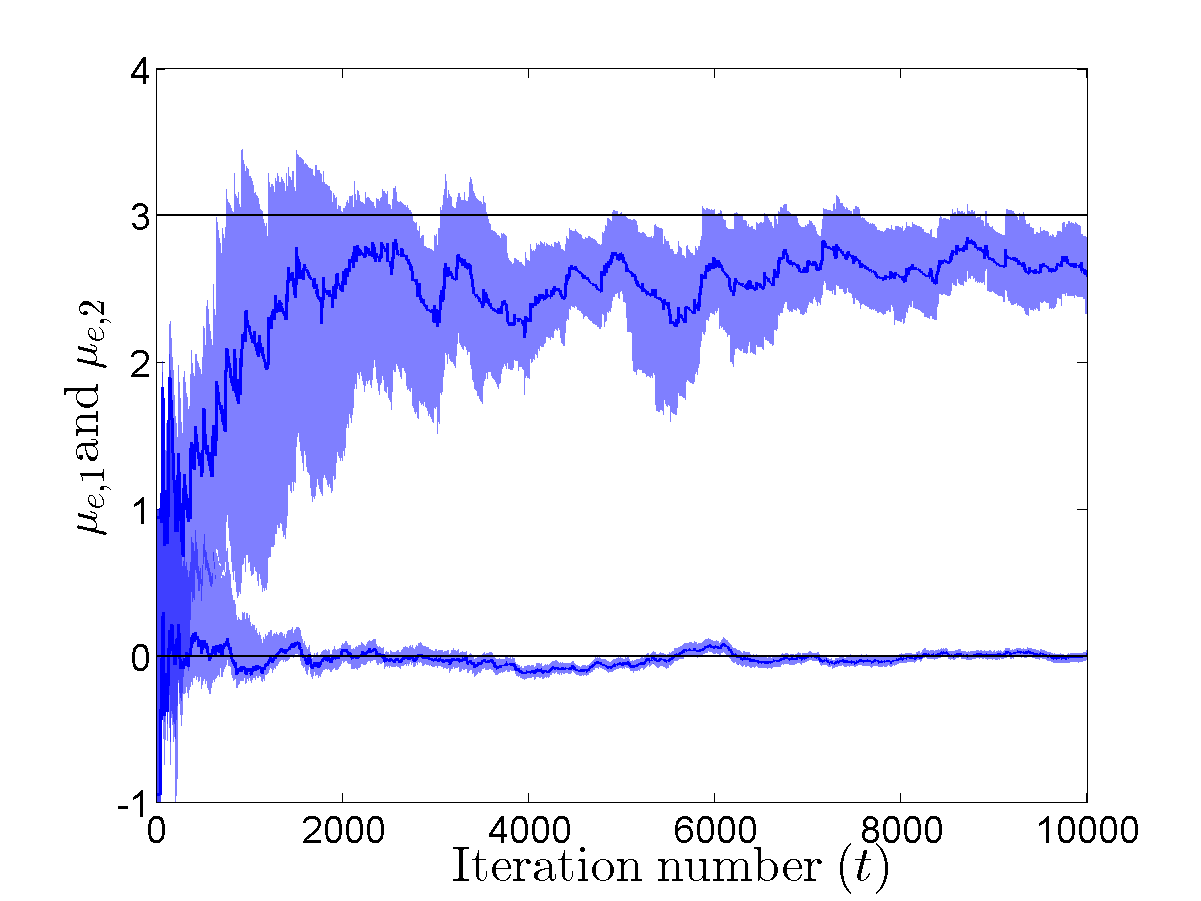}%
  \includegraphics[width=0.25\linewidth]{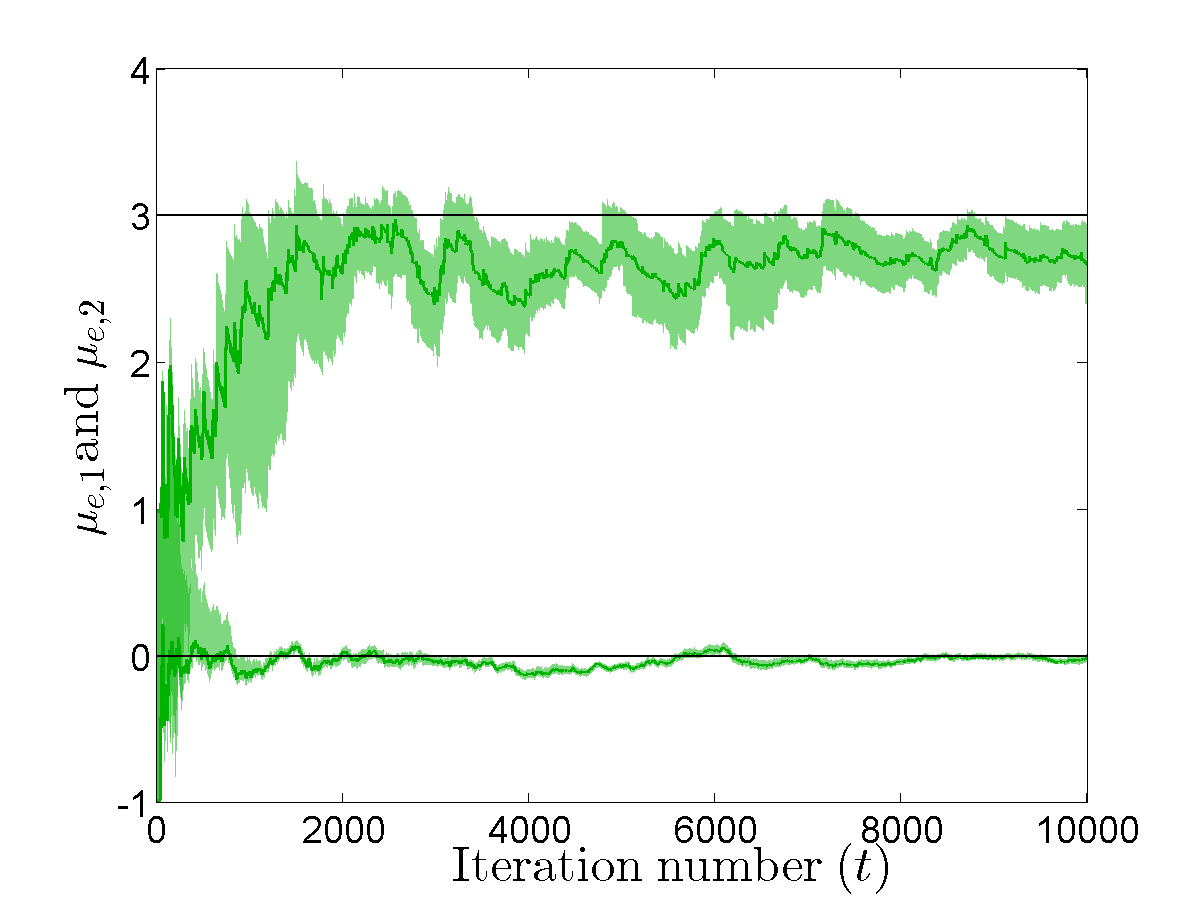}\\
  \includegraphics[width=0.25\linewidth]{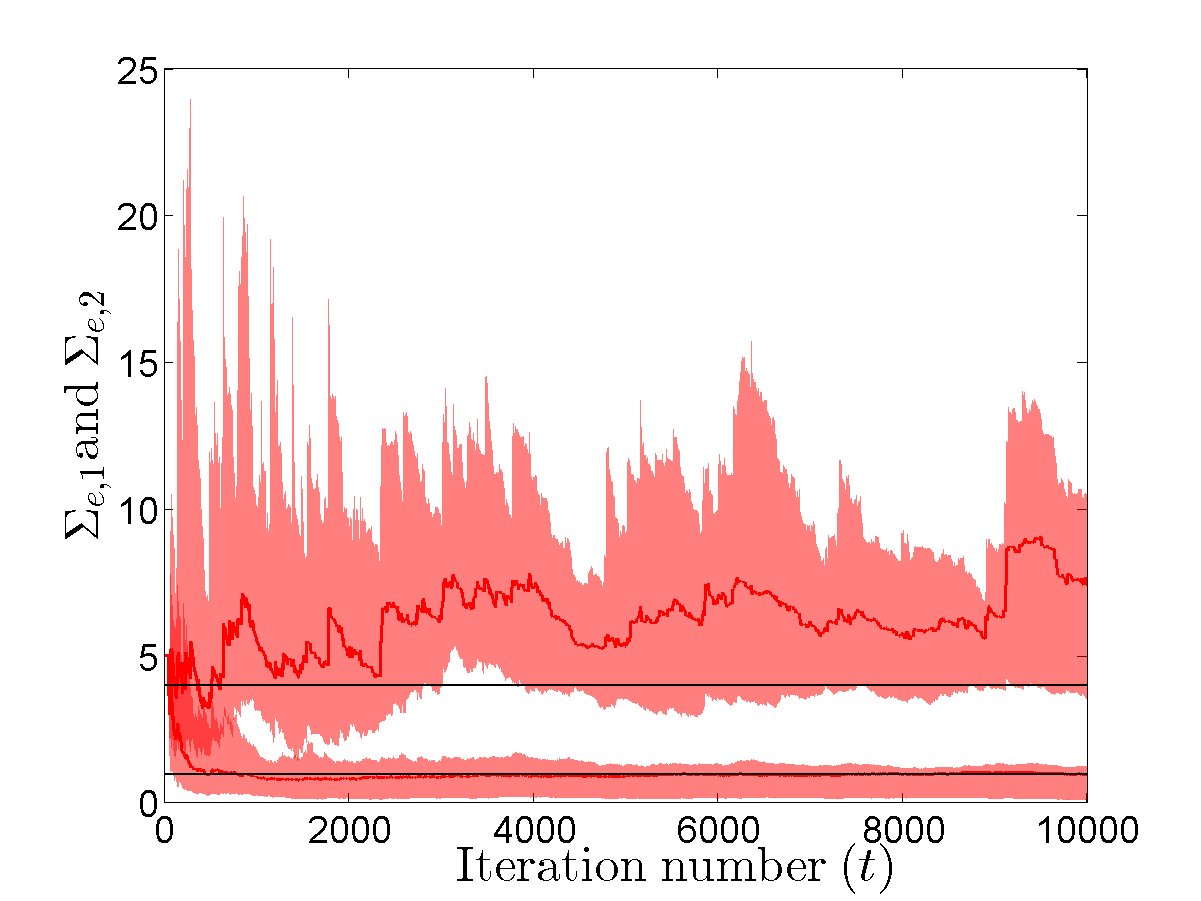}%
  \includegraphics[width=0.25\linewidth]{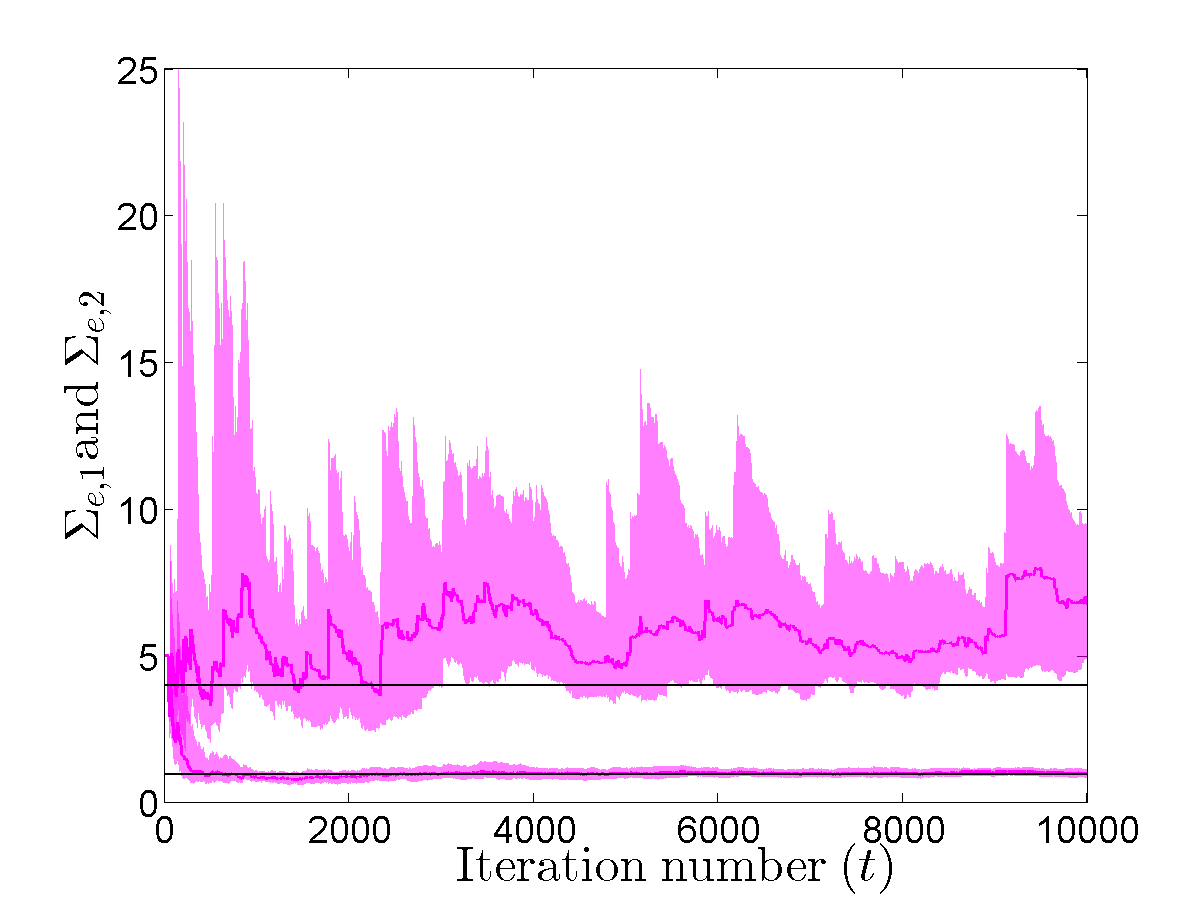}%
  \includegraphics[width=0.25\linewidth]{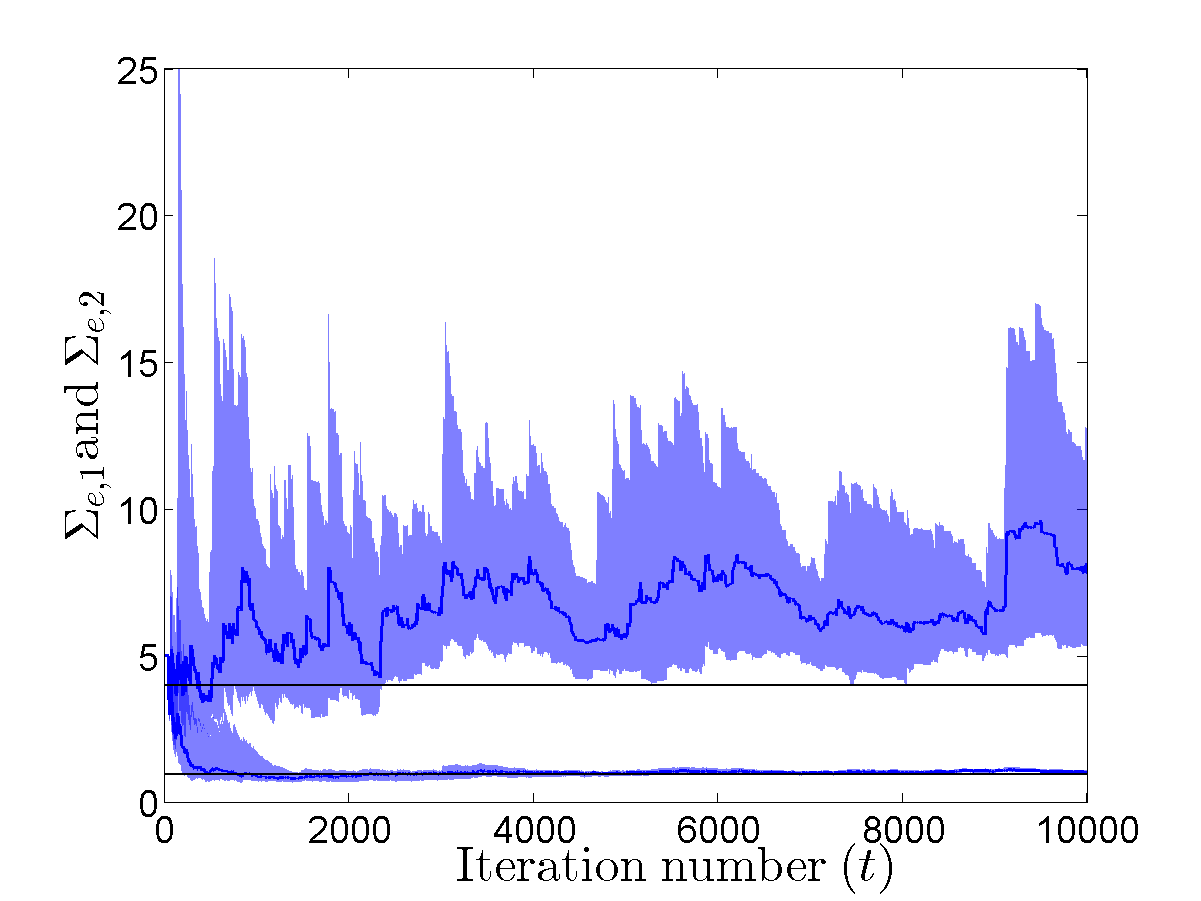}%
  \includegraphics[width=0.25\linewidth]{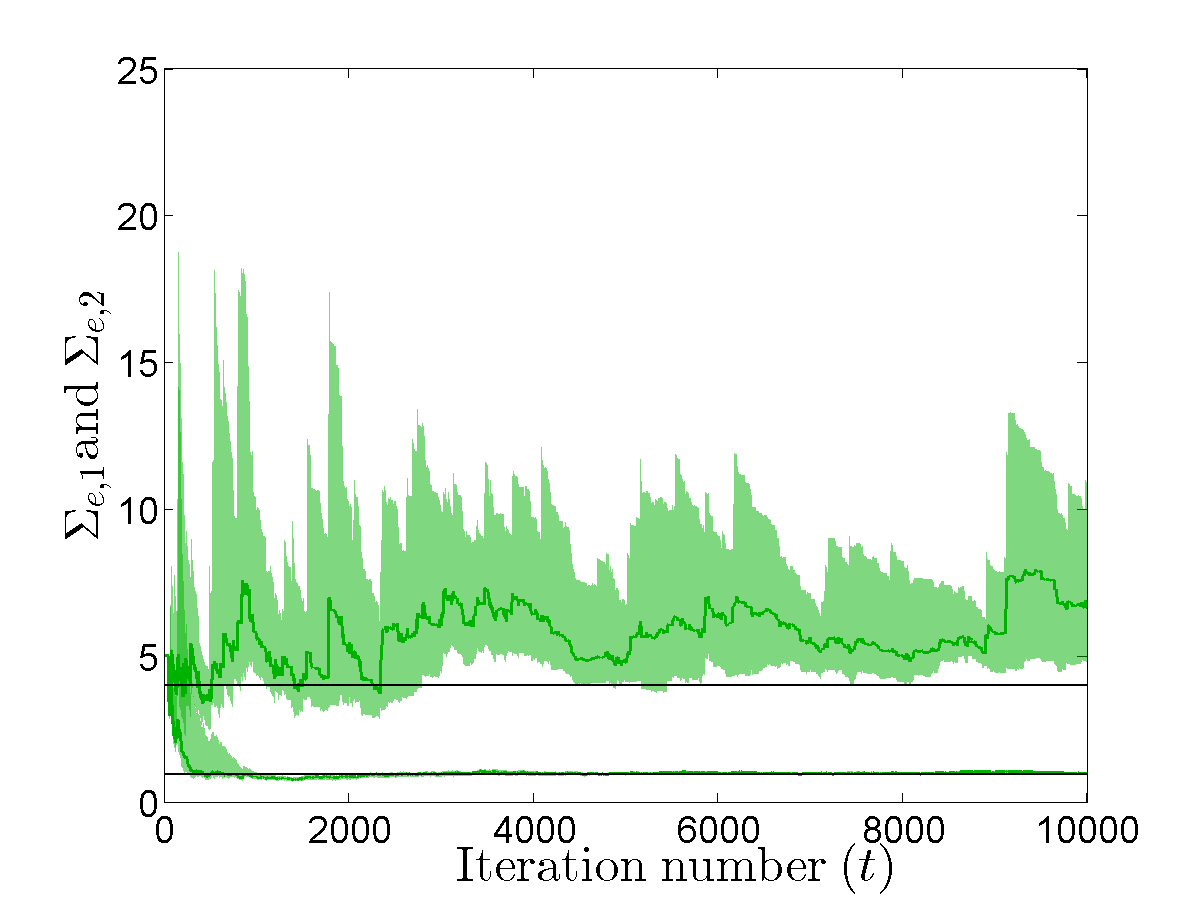}\\
  \includegraphics[width=0.25\linewidth]{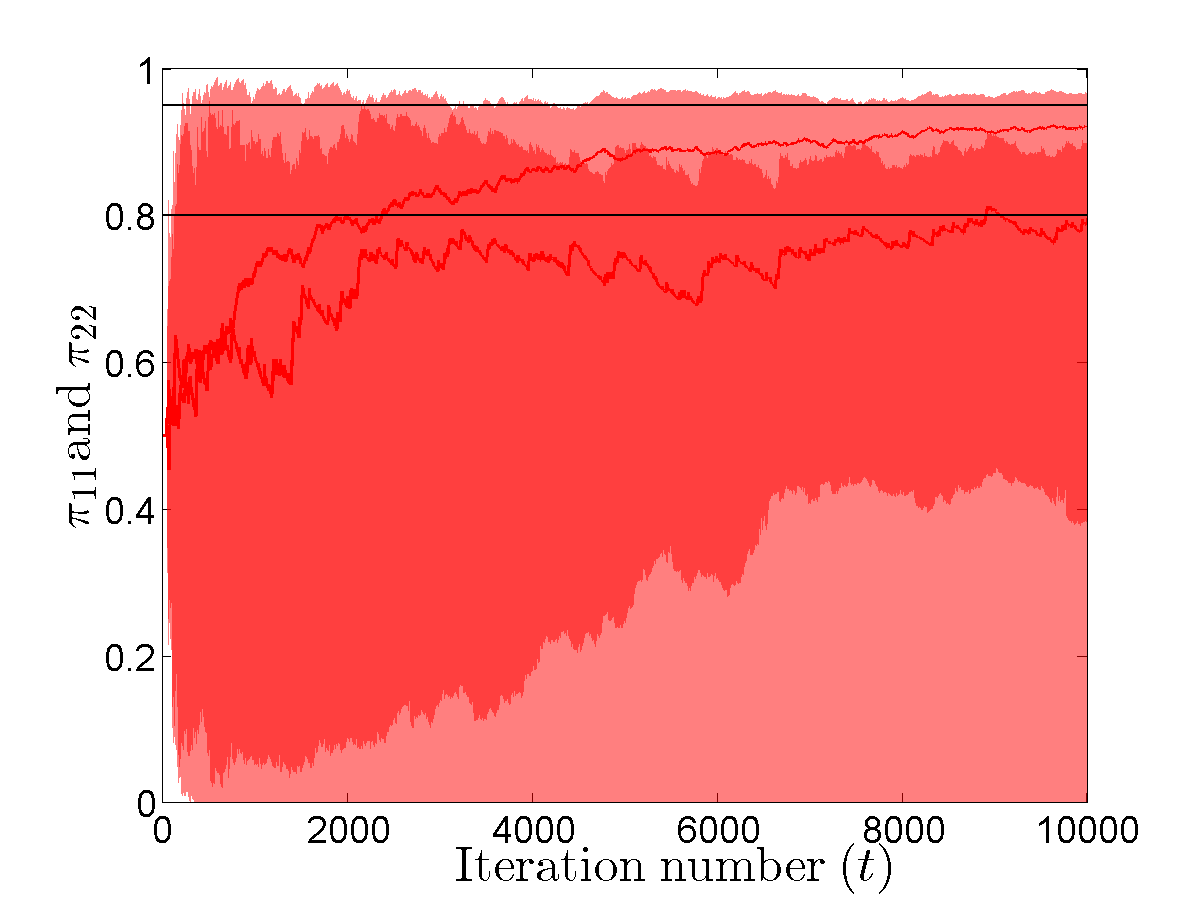}%
  \includegraphics[width=0.25\linewidth]{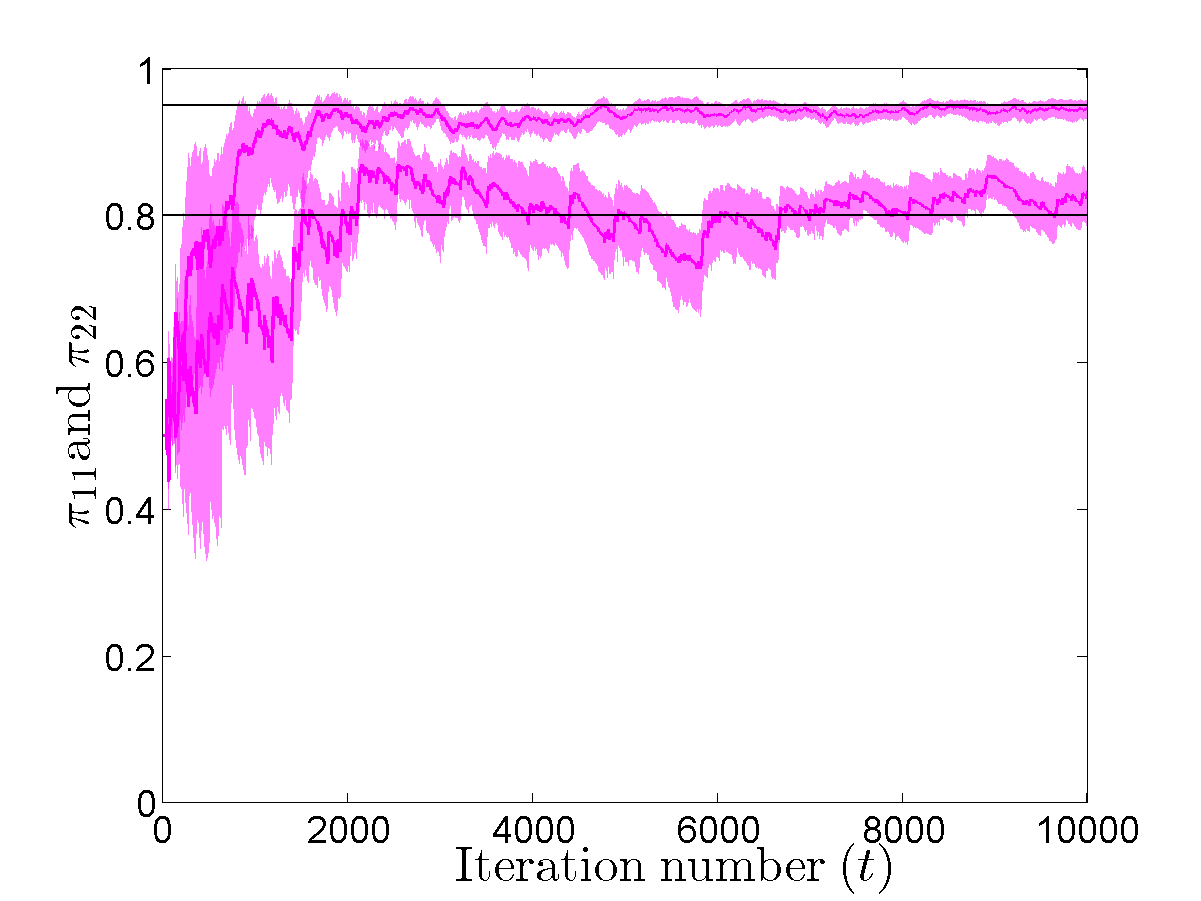}%
  \includegraphics[width=0.25\linewidth]{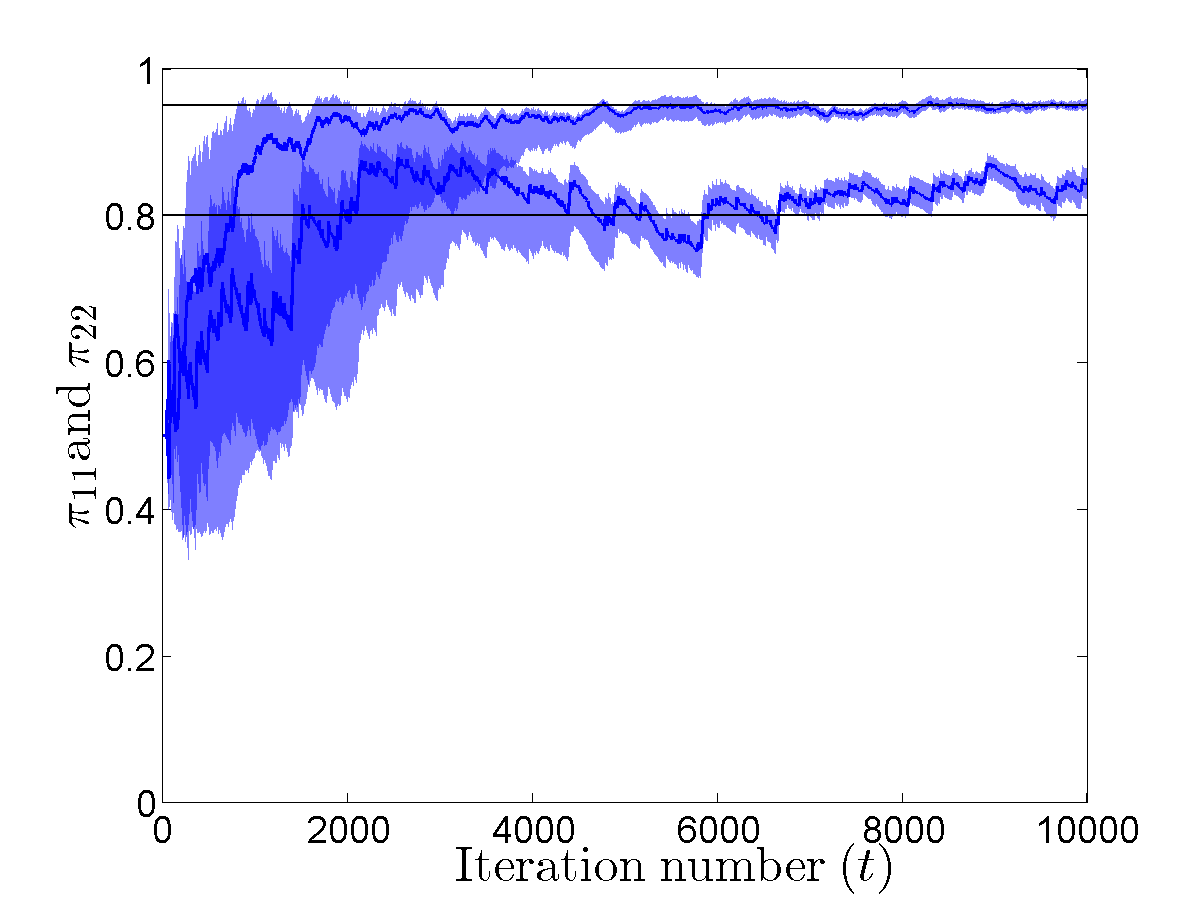}%
  \includegraphics[width=0.25\linewidth]{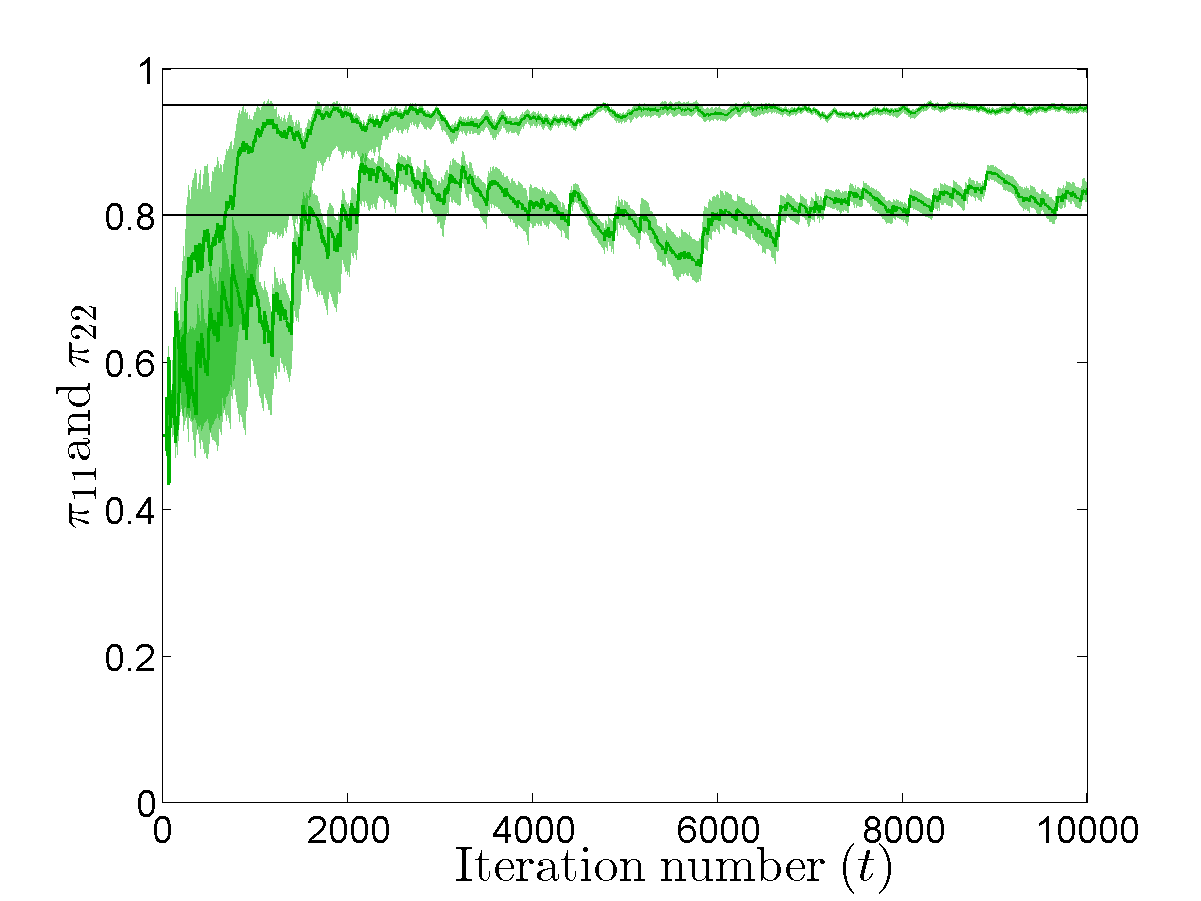}
  \caption{Estimation results over \thsnd{10}
  iterations. From left to right: \PFpath, \RBPFpath, \PFFS, and \RBPFFS. From top to bottom:
    $(\mu_{\textrm{e},1}, \mu_{\textrm{e},2})$, $(\Sigma_{\textrm{e},1}, \Sigma_{\textrm{e},2})$, and $(\pi_{11}, \pi_{22})$.
    The lines show the averages and the transparent shaded areas show the upper and lower bounds over 100 independent runs on the same data batch.}
  \label{fig:measurement_noise}
\end{figure*}

We compare the estimation performance of four different online EM algorithms:
\begin{itemize}
\item \textbf{\PFpath:} Path-based particle filter \cite{Cappe2009}.
\item \textbf{\PFFS:} Forward-smoothing-based particle filter \cite{DelMoral10}.
\item \textbf{\RBPFpath:} Path-based \rbpf (Algorithm~\ref{alg:rbpf_onlinem}).
\item \textbf{\RBPFFS:} Forward-smoothing-based \rbpf (Algorithm~\ref{alg:rbpf_onlinem}).
\end{itemize}
We simulate a batch of \thsnd{10} measurements $y_{1:n}$ and run all the algorithms 100 times
on the same data to investigate the errors arising from the Monte Carlo approximations.
All methods are bootstrap implementations with $\Np = 150$ particles and the step size sequence $\gamma_t = t^{-0.7}$.
See Appendix~\ref{app:noises} for further details on the implementation.
The results are shown in Figure~\ref{fig:measurement_noise}. Table~\ref{table:mcvar} reports the
time averaged Monte Carlo variances for the different methods.
The runtimes of the algorithms are given in Table~\ref{table:executiontimes}.
 All simulations are run in Matlab(R) R2012b on a standard laptop Intel(R)
Core(TM) i7-M640 2.80GHz platform with 8GB of RAM.

It can be seen that Rao-Blackwellization has a positive effect on reducing the Monte Carlo variance of the estimates. More specifically, the Monte Carlo
variances for \RBPFpath and \RBPFFS are smaller than for \PFpath and \PFFS, respectively.
Taking the computation times into account, \RBPFpath appears to provide a good trade-off between runtime and accuracy. The estimation performance
of \RBPFpath is similar to \RBPFFS and \PFFS, but with a significantly lower computational time.

\begin{table}[ht]
\begin{center}
  \caption{Time averaged Monte Carlo variances} % title of Table
  % used for centering table
  \begin{tabular}{l c c c c } % centered columns (4 columns)
    \hline
    & {\scriptsize\PFpath} & {\scriptsize\RBPFpath} & {\scriptsize\PFFS} & {\scriptsize\RBPFFS}   \\ [0.5ex] % inserts table
    % heading
    \hline\hline
$\mu_{\textrm{e},1}$ $(\times 10^{-3})$ & 21.8 &   4.16 &   3.46 &   1.23 \\
$\mu_{\textrm{e},2}$ $(\times 10^{-2})$ & 38.0 &   3.98 &   4.45 &   1.71 \\
$\Sigma_{\textrm{e},1}$ $(\times 10^{-2})$ &  7.66 &   1.77 &   2.05 &   1.54 \\
$\Sigma_{\textrm{e},2}$ $(\times 10^{0})$ &  2.42 &   1.74 &   1.87 &   1.32 \\
$\pi_{11}$ $(\times 10^{-4})$ &343 &   6.10 &   8.10 &   1.53 \\
$\pi_{22}$ $(\times 10^{-4})$ &268&   3.17 &  12.6 &   1.62 \\
    [1ex]      % [1ex] adds vertical space
    \hline %inserts single line
  \end{tabular}
\label{table:mcvar} % is used to refer this table in the text
\end{center}
\end{table}

\begin{table}[ht]
\begin{center}
  \caption{Average runtimes in milliseconds/time step} % title of Table
  % used for centering table
  \begin{tabular}{c c c c } % centered columns (4 columns)
    \hline
    {\scriptsize\PFpath} & {\scriptsize\RBPFpath} & {\scriptsize\PFFS} & {\scriptsize\RBPFFS}   \\ [0.5ex] % inserts table
    %\PFpath & \RBPFpath & \PFFS& \RBPFFS   \\ [0.5ex] % inserts table
    % heading
    \hline\hline
    0.42 & 1.08 & 2.90 & 6.71  \\ % inserting body of the table
    [1ex]      % [1ex] adds vertical space
    \hline %inserts single line
  \end{tabular}
\label{table:executiontimes} % is used to refer this table in the text
\end{center}
\end{table}

\subsection{Transition Probability Estimation for \jmls} %Jump Markov Linear Systems}
In this section, we illustrate the performance of the algorithm on a jump Markov linear model. JMLS are studied thoroughly in the literature and many dedicated algorithms are proposed for the estimation of the transition probabilities  which exploit the linear Gaussian structure in the model \cite{Jilkov2004,Orguner2006online,Orguner2008}. Using inaccurate TPMs may lead to performance degradation of the state estimation, due to the sensitivity of the multiple model state estimators to the TPM used. The uncertainty regarding the TPM is a major issue in the application of multiple models to real-life problems \cite{Jilkov2004}. The proposed online EM solution is naturally applicable to linear systems and does not involve any IMM filtering type approximations. In IMM type mixing approximations, many components in the posterior are systematically collapsed into a single Gaussian which deteriorates the statistics associated with the dominant modes and degrades the performance. In the simulation below, we have considered the benchmark model originally given in \cite{Jilkov2004}, and used in \cite{Orguner2006online} and \cite{Orguner2008} for comparison of different TPM estimation algorithms:
\begin{align}
x_{t+1}&=x_t+v_t\\
y_t&=r_t x_t+ (100- 90r_t)e_t
\end{align}
where $x_0\sim\mathcal{N}(x_0;0,20^2)$, $v_t\sim\mathcal{N}(v_t;0,2^2)$, and $e_t\sim\mathcal{N}(e_t;0,1)$ with $x_0,v_t$ and $e_t$ being mutually
independent for $t=1,\,2,\,\ldots$. The mode sequence $r_t \in \{0,1\}$ is a 2-state homogenous Markov process with TPM given as,
\begin{align}
\Pi=\left[
                                                                                                                          \begin{array}{cc}
                                                                                                                            0.6 & 0.4 \\
                                                                                                                            0.85 & 0.15 \\
                                                                                                                          \end{array}
                                                                                                                        \right].
\end{align}
 This system corresponds to a system with frequent measurement failures with the modal state $r_t=0$ corresponding to a failure. The online EM algorithm is run on the simulated measurements of this system with initial transition probabilities $\widehat{\pi}_{00}^{(0)}=\widehat{\pi}_{11}^{(0)}=0.5$. We compare three algorithms in a single run using the same data set used in \cite{Orguner2008}.  The first algorithm is a Kullback-Leibler-distance-based TPM estimation method, denoted as IMM-KL, which is proposed in \cite{Orguner2006online}. The second algorithm is a maximum-likelihood-based method, denoted as IMM-ML, which is presented in \cite{Orguner2008}. These two algorithms rely on the aforementioned IMM approximations. The third algorithm is the proposed \RBPFpath method,
using 500 particles and the step size sequence $\gamma_t = t^{-0.95}$.
\begin{figure}[]
	\begin{center}
	\includegraphics[width=0.6\columnwidth]{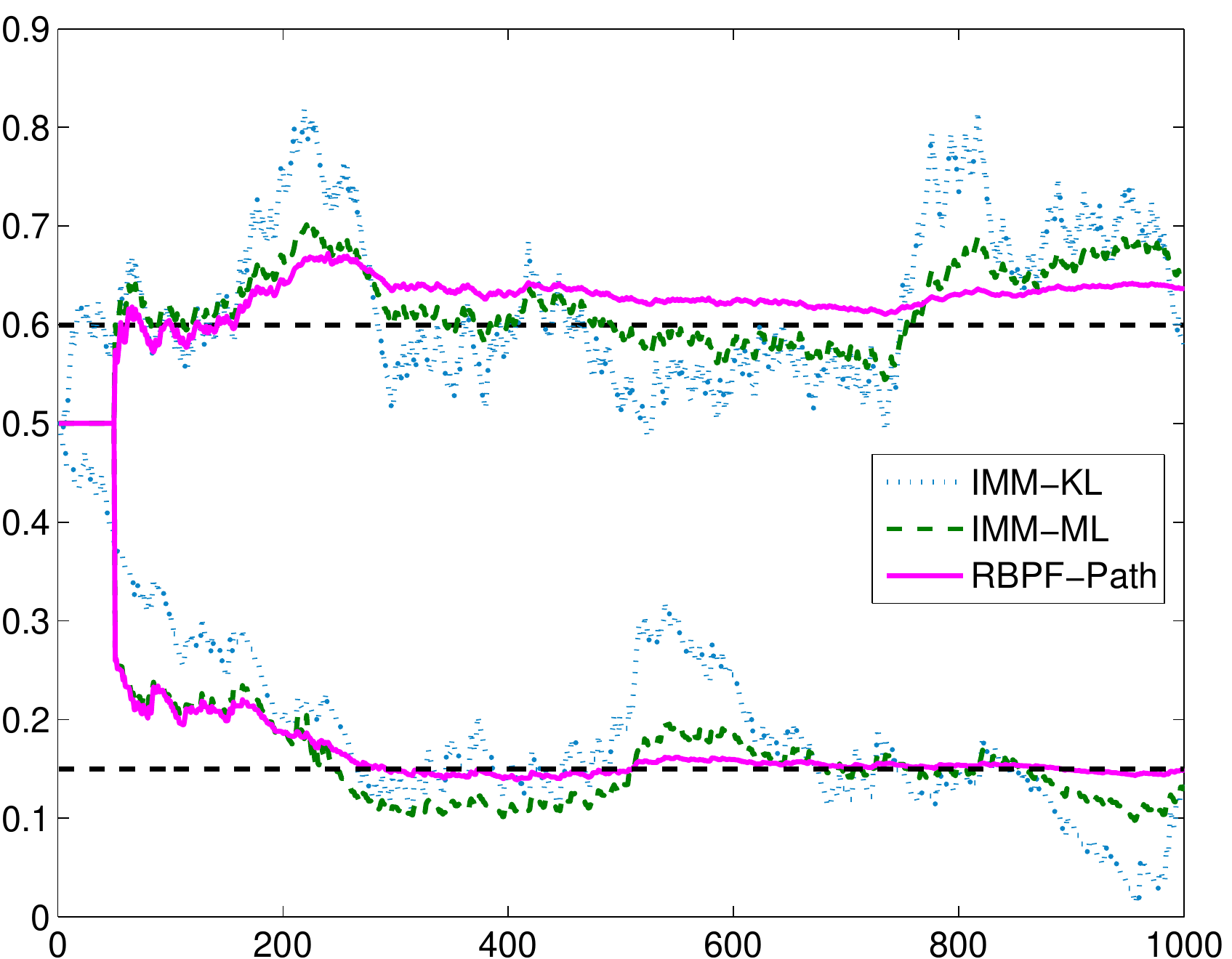}
	\caption{Transition probability matrix estimation performance. The estimated transition probabilities are plotted for IMM-KL (dashed blue line),
          IMM-ML (green dotted line) and \RBPFpath (solid magenta line).}
	\label{fig: TPM}
	\end{center}
\end{figure}
In Figure~\ref{fig: TPM}, the estimated transition probabilities of the three algorithms are depicted.
The \RBPFpath appears to provide satisfactory results, showing fast convergence to the vicinity of the true parameters at the beginning and providing smoother estimates towards the end. It is worth to note that, contrary to the special purpose algorithms IMM-KL and IMM-ML,
 this is accomplished without exploiting the linearity of the dynamic modes.

\subsection{Mobile Terminal Positioning in Wireless Networks}
The results in this section are provided to illustrate the validity of the proposed method on real data. We consider the mobile terminal (MT) positioning example in a wireless network, where time of arrival (ToA) measurements from three base stations are available to determine the position of the MT. The measurements have been collected during a field trial performed in Kista, Sweden; see \cite{Medbo2009} for more details on experimental setup. The scenario can be considered as dense urban, where many multi-storey buildings prevent that the radio signal from the base stations (BSs) arrive via the direct line-of-sight ($\LOS$) path at the MT. Due to multiple reflections from buildings, the radio signals often propagate via an indirect non-line-of-sight($\NLOS$) path to the MT. In the literature, these switching propagation conditions are often modeled with a two-state Markov chain affecting the noise distribution of the measurement; see for instance \cite{Fritsche2009a,Liao2006}. This approach is also followed here, but with the assumption that the underlying measurement noise statistics as well as the parameters of the Markov chain are unknown and have to be estimated.

It is assumed that the MT motion can be modeled with a nearly constant velocity model, according to
\begin{align}
x_{t}=F x_{t-1}+\Gamma v_{t}
\label{MTmotion1}
\end{align}
with
\begin{align}
F=\left[\begin{array}{cccc}
1 & \Delta T & 0 & 0 \\
0 & 1 & 0 & 0 \\
0 & 0 & 1 & \Delta T \\
0 & 0 & 0 & 1 \\
\end{array}\right],\quad \Gamma=\left[\begin{array}{cc}
\Delta T^2/2 & 0\\
\Delta T & 0 \\
0 & \Delta T^2/2\\
0 & \Delta T\\
\end{array}\right],
\label{MTmotion2}
\end{align}
where $x_{t}=[x_{\textrm{MT},t},\dot{x}_{\textrm{MT},t},y_{\textrm{MT},t},\dot{y}_{\textrm{MT},t}]^\+$ is the MT position and velocity vector and $\Delta T=0.53$ s is the sampling time. The noise is distributed according to $v_{t}\sim\mathcal{N}(0,\Sigma_{\textrm{v}})$ with $\Sigma_{\textrm{v}}=\sigma_v^2 I_2$, where $I_2$ denotes the $2\times2$ identity matrix. The switching is modeled with a 2-state Markov chain $\{r_t\}$, where the state $r_t = 1$ is assigned to the event $\LOS$ and the state $r_t = 2$ is assigned to the event $\NLOS$. The Markov chain is assumed to be time-homogeneous with transition probability matrix $\Pi$. In the following, the ToA measurements are expressed in terms of distance measurements (by multiplication with speed of light), so that the measurement at each BS can be described with
\begin{align}
y_{t}=h(x_{t})+e_t^{(r_t)},
\label{MTmeasurment1}
\end{align}
where
\begin{align}
h(x_{t})=\sqrt{(x_{\textrm{MT},t}-x_{\textrm{BS},t})^2+(y_{\textrm{MT},t}-y_{\textrm{BS},t})^2}
\label{MTmeasurment2}
\end{align}
and $(x_{\textrm{BS},t},y_{\textrm{BS},t})$ are the BS position coordinates. The noise is distributed according to $e_t^{(r_t)}\sim\mathcal{N}(\mu_{{r_t}},\sigma^2_{{r_t}})$, meaning that the LOS and NLOS errors are modeled with a Gaussian distribution with different means and variances according to
\begin{align}
  \mu_{{r_t}} &=
  \begin{cases}
    \mu_{\textrm{LOS}}, & r_t = 1,\\
    \mu_{\textrm{NLOS}}, & r_t = 2,
  \end{cases} &
  \sigma^2_{{r_t}} &=
  \begin{cases}
    \sigma^2_{\textrm{LOS}}, & r_t = 1,\\
    \sigma^2_{\textrm{NLOS}}, & r_t = 2.
  \end{cases}
\label{MTmeasurment3}
\end{align}
For simplicity,
only BS 3, which is severely affected by switching propagation conditions, is modeled according to (\ref{MTmeasurment1}). The measurement noise of the other two BSs is assumed to be Gaussian distributed, where the mean and variance have been determined prior to running the algorithm. Thus, the unknowns stemming from the measurement model of BS 3 can be collected in $\theta = \left(\mu_{\textrm{LOS}},\mu_{\textrm{NLOS}},\sigma^2_{\textrm{LOS}},\sigma^2_{\textrm{NLOS}}, \Pi \right)$.

\begin{figure}[]
	\begin{center}
	\includegraphics[width=0.6\columnwidth]{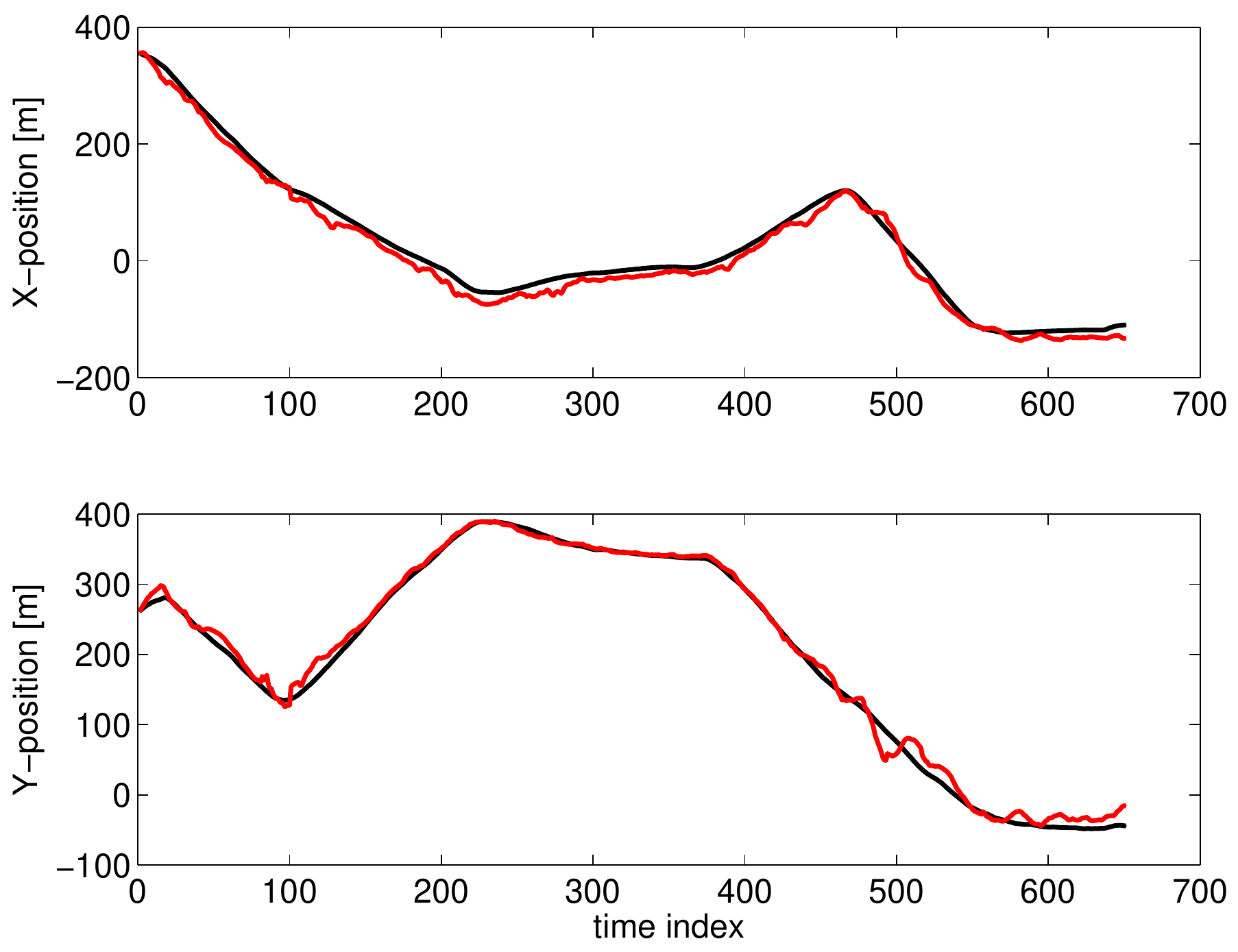}
	\caption{Estimated (red line) and true position (black line) coordinates of the moving agent.}
	\label{fig:BS_Position}
	\end{center}
\end{figure}

\begin{figure}[]
	\begin{center}
	\includegraphics[width=0.6\columnwidth]{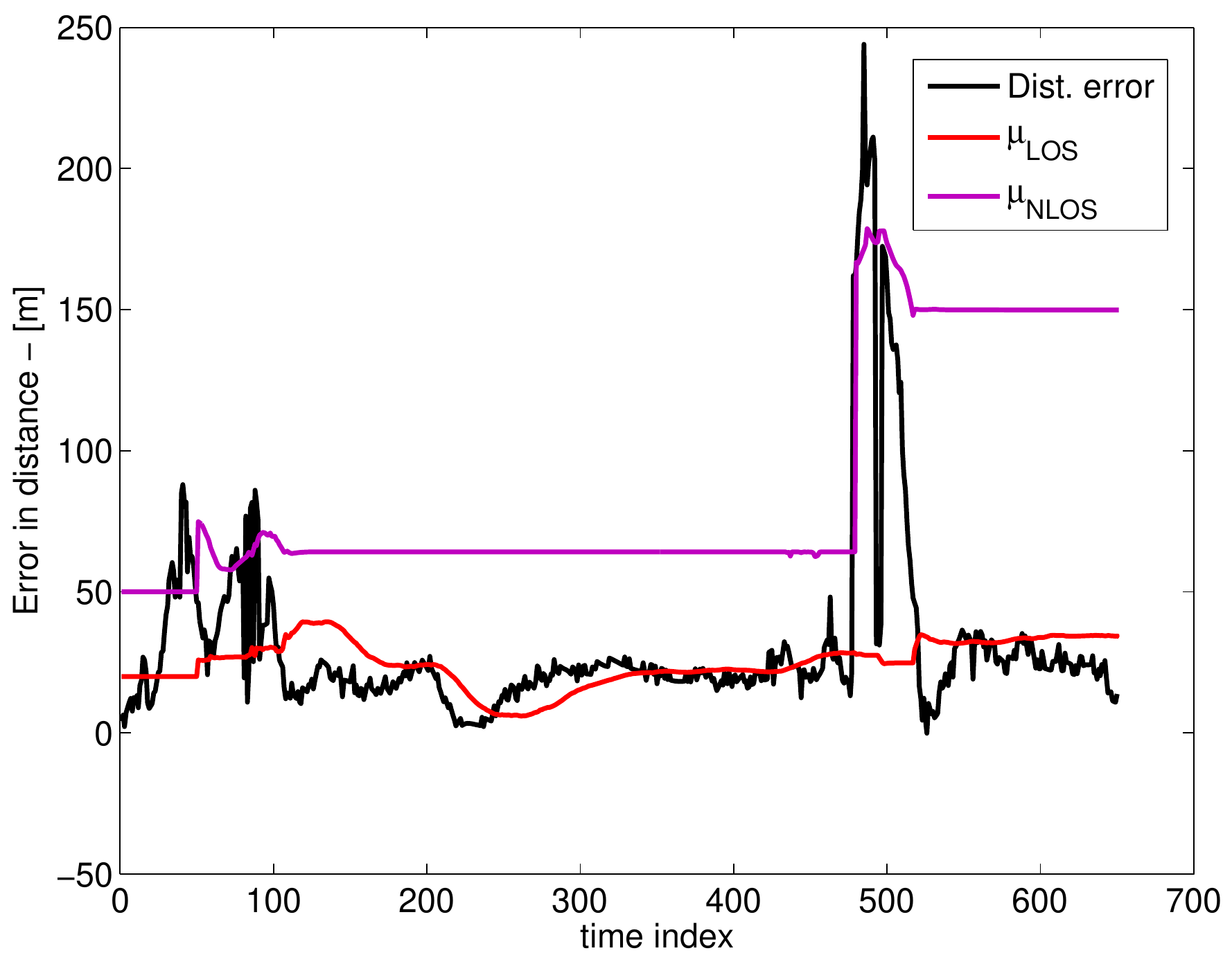}
	\caption{Estimated mean terms $\mu_{LOS}, \mu_{NLOS}$ plotted on error in distance measurements of BS 3, calculated from the distance measured by GPS.}
	\label{fig:BS_Biases}
	\end{center}
\end{figure}

\begin{figure}[]
	\begin{center}
	\includegraphics[width=0.6\columnwidth]{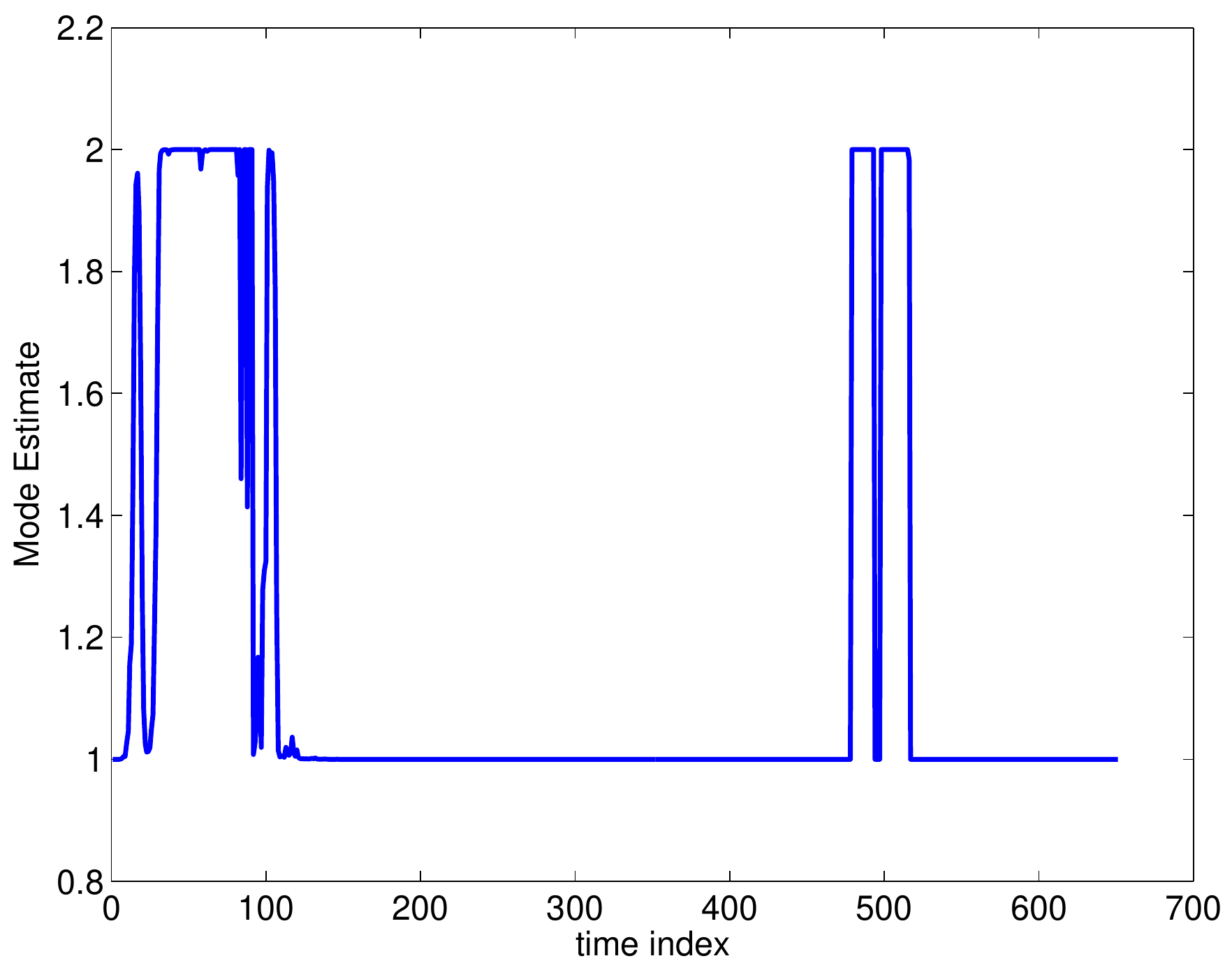}
	\caption{Mode estimate \vs time where '1' is for the LOS mode and '2' is for the NLOS mode.}
	\label{fig:BS_ModeProb}
	\end{center}
\end{figure}

In Figure~\ref{fig:BS_Position}, the estimation results for the MT coordinates are shown using the proposed RBPF-Path algorithm with 500 particles. The variances $\sigma^2_{\textrm{LOS}},\sigma^2_{\textrm{NLOS}}$ are constrained to be less than 100.  It can be observed that the estimated MT position coordinates follow the true ones. The mean terms $\mu_{\textrm{LOS}}$ and $\mu_{\textrm{NLOS}}$ together with the error in distance measurements of BS 3, obtained from the true distance measured by GPS is shown in Figure~\ref{fig:BS_Biases}. It can be observed, that the (time varying) biases in the distance measurements can be generally well tracked by the algorithm. In Figure~\ref{fig:BS_ModeProb}, the mode estimates of the algorithm is plotted. The switchings from $\LOS$ to $\NLOS$ modes are tracked successfully.

\section{Conclusion}
We have proposed a method based on the online EM algorithm for joint state estimation and identification of \jmnls.
We use Rao-Blackwellization to exploit the structure of \jmnls, resulting in a significant improvement in the
estimation accuracy. The algorithm is applicable to a large class of models which involve sudden regime changes,
unknown parameters and heavy non-linearities as illustrated via simulations.
The algorithm was also successfully tested on real data for localization in a wireless network.

\appendix[Sufficient Statistics for Noise Parameters in Jump Markov Gaussian Systems]\label{app:noises}
Consider the jump Markov System given below.
\begin{subequations}
\label{eq:JMGS}
\begin{align}
x_t&=f_{r_t}(x_{t-1})+v_t^{(r_t)},\label{eq:JMGS_PM}\\
y_t&=h_{r_t}(x_t)+e_t^{(r_t)}\label{eq:JMGS_MM},
\end{align}
\end{subequations}
where the noise is distributed according to $v_t^{(r_t)}\sim\mathcal{N}(\mu_{\textrm{v},r_t},\Sigma_{\textrm{v},r_t})$ and
$e_t^{(r_t)}\sim\mathcal{N}(\mu_{\textrm{e},r_t},\Sigma_{\textrm{e},r_t})$.
The unknowns parameters are $\theta = \left(\{\theta_k\}_{k=1}^K, \Pi \right)$, where $\Pi$ refers to the $K\times K$ transition matrix with entries $[\Pi]_{k\ell} = \pi_{k\ell}$ and  $\theta_{k}=\{\mu_{\textrm{v},k},\Sigma_{\textrm{v},k},\mu_{\textrm{e},k},\Sigma_{\textrm{e},k}\}$. Below, we provide the sufficient statistics $s_{t}(\xi_{t}, \xi_{t-1})$ as well as the closed-form expressions for the mappings $\Lambda_k(\cdot)$ appearing in \eqref{eq:EM_JMNLS:parameterupdate}. These mappings can be found by explicitly evaluating the M-step. Similarly to \cite{Cappe2011}, for jump Markov Gaussian systems, the parameters can be updated according to:
\begin{subequations}
  \begin{align}
        \widehat \pi_{k\ell} &= \frac{\P{k\ell}{t} }{ \sum_{j=1}^K \P{kj}{t} },\\
        \widehat \mu_{\textrm{v},k} &=\frac{\S{k}{\textrm{v}}\langle 1 \rangle}{ \mathcal{S}^{(2)}_{k}},&
        \widehat \Sigma_{\textrm{v},k} &=\frac{\S{k}{\textrm{v}}\langle 2 \rangle}{ \mathcal{S}^{(2)}_{k}}-\widehat \mu_{\textrm{v},k} \widehat \mu_{\textrm{v},k}^\+,\\
        \widehat \mu_{\textrm{e},k} &= \frac{\S{k}{\textrm{e}}\langle 1 \rangle}{ \mathcal{S}^{(2)}_{k}},&
        \widehat \Sigma_{\textrm{e},k} &= \frac{\S{k}{\textrm{e}}\langle 2 \rangle}{ \mathcal{S}^{(2)}_{k}}-\widehat \mu_{\textrm{e},k} \widehat \mu_{\textrm{e},k}^\+,
  \end{align}
\end{subequations}

The corresponding sufficient statistics are given by
\begin{subequations}
\begin{align}
s_{k\ell,t}^{\langle 1 \rangle}&=\I{r_{t}=k,r_{t-1}=l},\\
s_{k,t}^{(2)}&=\I{r_{t}=k},\\
s_{k,\textrm{v},t}^{(3)}\langle 1 \rangle&=\I{r_{t}=k}[x_{t}-f_{r_{t}}(x_{t-1})],\\
s_{k,\textrm{v},t}^{(3)}\langle 2 \rangle&=\I{r_{t}=k}[x_{t}-f_{r_{t}}(x_{t-1})][\cdot]^\+,\\
s_{k,\textrm{e},t}^{(3)}\langle 1 \rangle&=\I{r_{t}=k}[y_{t}-h_{r_{t}}(x_{t})]\\
s_{k,\textrm{e},t}^{(3)}\langle 2 \rangle&=\I{r_{t}=k}[y_{t}-h_{r_{t}}(x_{t})][\cdot]^\+.
\end{align}
\end{subequations}

\section*{Acknowledgment}
The authors gratefully acknowledge the financial support from the Swedish Research Council under the Linnaeus Center (CADICS) and would like to thank Umut Orguner for providing the code of his previous works for comparison in Section 6.2.

\bibliographystyle{plain}
\bibliography{arxiv}

\begin{thebibliography}{10}

\bibitem{Andrieu03JM}
C.~Andrieu, M.~Davy, and A.~Doucet.
\newblock Efficient particle filtering for jump {M}arkov systems. application
  to time-varying autoregressions.
\newblock {\em Signal Processing, IEEE Transactions on}, 51(7):1762--1770,
  2003.

\bibitem{blom1988}
HAP Blom and Y.~Bar-Shalom.
\newblock The interacting multiple model algorithm for systems with {M}arkovian
  switching coefficients.
\newblock {\em Automatic Control, IEEE Transactions on}, 33(8):780--783, 1988.

\bibitem{Cappe2009}
O.~Capp\'e.
\newblock Online sequential {M}onte {C}arlo {EM} algorithm.
\newblock In {\em Statistical Signal Processing, 2009. SSP '09. IEEE/SP 15th
  Workshop on}, pages 37 --40, Sept. 2009.

\bibitem{Cappe2011}
O.~Capp\'e.
\newblock Online {EM} algorithm for hidden {M}arkov models.
\newblock {\em Journal of Computational and Graphical Statistics},
  20(3):728--749, 2011.

\bibitem{CappeMR:2005}
O.~Capp\'e, E.~Moulines, and T.~Ryd\'en.
\newblock {\em Inference in Hidden {M}arkov Models}.
\newblock Springer, 2005.

\bibitem{Cappe2005}
O.~Capp\'e, E.~Moulines, and T.~Ryd\'en.
\newblock {\em Inference in Hidden {M}arkov Models}.
\newblock Springer Series in Statistics. Springer Science + Business Media,
  LLC, New York, NY, USA, 2005.

\bibitem{Caron2007}
F.~Caron, M.~Davy, E.~Duflos, and P.~Vanheeghe.
\newblock Particle filtering for multisensor data fusion with switching
  observation models: Application to land vehicle positioning.
\newblock {\em Signal Processing, IEEE Transactions on}, 55(6):2703--2719,
  2007.

\bibitem{CarvalhoL:2007}
C.~M. Carvalho and H.~F. Lopes.
\newblock Simulation-based sequential analysis of {M}arkov switching stochastic
  volatility models.
\newblock {\em Computational Statistics \& Data Analysis}, 51:4526--4542, 2007.

\bibitem{ChenL:2000}
R.~Chen and J.~S. Liu.
\newblock Mixture {K}alman filters.
\newblock {\em Journal of the Royal Statistical Society: Series {B}},
  62(3):493--508, 2000.

\bibitem{Chopin:2004}
N.~Chopin.
\newblock Central limit theorem for sequential {M}onte {C}arlo methods and its
  application to {B}ayesian inference.
\newblock {\em The Annals of Statistics}, 32(6):2385--2411, 2004.

\bibitem{cinquemani2007general}
E.~Cinquemani, R.~Porreca, G.~Ferrari-Trecate, and John Lygeros.
\newblock A general framework for the identification of jump {M}arkov linear
  systems.
\newblock In {\em Decision and Control, 2007 46th IEEE Conference on}, pages
  5737--5742. IEEE, 2007.

\bibitem{DelMoral10}
P.~Del~Moral, A.~Doucet, and S.~Singh.
\newblock Forward smoothing using sequential {M}onte {C}arlo.
\newblock {\em arXiv preprint arXiv:1012.5390}, 2010.

\bibitem{Dempster1977}
A.~P. Dempster, N.~M. Laird, and D.~B. Rubin.
\newblock Maximum likelihood from incomplete data via the {EM} algorithm.
\newblock {\em Journal of the Royal Statistical Society. Series B
  (Methodological)}, 39(1):1--38, 1977.

\bibitem{DoucetGA:2000}
A.~Doucet, S.~J. Godsill, and C.~Andrieu.
\newblock On sequential {M}onte {C}arlo sampling methods for {B}ayesian
  filtering.
\newblock {\em Statistics and Computing}, 10(3):197--208, 2000.

\bibitem{Doucet2001a}
A.~Doucet, N.~J. Gordon, and V.~Krishnamurthy.
\newblock Particle filters for state estimation of jump {M}arkov linear
  systems.
\newblock {\em {IEEE} Transactions on Signal Processing}, 49(3):613--624, 2001.

\bibitem{DoucetJ:2011}
A.~Doucet and A.~Johansen.
\newblock A tutorial on particle filtering and smoothing: Fifteen years later.
\newblock In D.~Crisan and B.~Rozovskii, editors, {\em The Oxford Handbook of
  Nonlinear Filtering}. Oxford University Press, 2011.

\bibitem{driessen2005}
H.~Driessen and Y.~Boers.
\newblock Efficient particle filter for jump {M}arkov nonlinear systems.
\newblock {\em IEE Proceedings-Radar, Sonar and Navigation}, 152(5):323--326,
  2005.

\bibitem{Fritsche2009a}
C.~Fritsche, U.~Hammes, A.~Klein, and A.~Zoubir.
\newblock Robust mobile terminal tracking in {NLOS} environments using
  interacting multiple model algorithm.
\newblock In {\em Proc. of IEEE International Conference on Acoustics, Speech
  and Signal Processing}, pages 3049--3052, Taipei, Taiwan, Apr. 2009.

\bibitem{carsten2012}
C.~Fritsche, E.~Ozkan, and F.~Gustafsson.
\newblock Online {EM} algorithm for jump {M}arkov systems.
\newblock In {\em Information Fusion (FUSION), 2012 15th International
  Conference on}, pages 1941--1946. IEEE, 2012.

\bibitem{GilksB:2001}
W.~R. Gilks and C.~Berzuini.
\newblock Following a moving target -- {M}onte {C}arlo inference for dynamic
  {B}ayesian models.
\newblock {\em Journal of the Royal Statistical Society. {S}eries {B}
  (Statistical Methodology)}, 63(1):127--146, 2001.

\bibitem{Gustafsson:2010a}
F.~Gustafsson.
\newblock Particle filter theory and practice with positioning applications.
\newblock {\em {IEEE} Aerospace and Electronic Systems Magazine}, 25(7):53--82,
  2010.

\bibitem{Jilkov2004}
V.P. Jilkov and X.R. Li.
\newblock Online {B}ayesian estimation of transition probabilities for
  {M}arkovian jump systems.
\newblock {\em Signal Processing, IEEE Transactions on}, 52(6):1620 -- 1630,
  june 2004.

\bibitem{Liao2006}
L.~Jung-Fen and C.~Bor-Sen.
\newblock Robust mobile location estimator with {NLOS} mitigation using {IMM}
  algorithm.
\newblock {\em Wireless Communications, IEEE Transactions on},
  5(11):3002--3006, Nov. 2006.

\bibitem{LeCorff2011}
S.~Le~Corff, G.~Fort, and E.~Moulines.
\newblock Online {E}xpectation {M}aximization algorithm to solve the {SLAM}
  problem.
\newblock In {\em Statistical Signal Processing Workshop (SSP), 2011 IEEE},
  pages 225 --228, Nice, France, June 2011.

\bibitem{li2003survey}
X.~R. Li and V.~P. Jilkov.
\newblock A survey of maneuvering target tracking part v: Multiple-model
  methods.
\newblock In {\em Conference on Signal and Data Processing of Small Targets},
  volume 4473, pages 559--581, 2003.

\bibitem{LindstenS:2013}
F.~Lindsten and T.~B. Sch\"on.
\newblock Backward simulation methods for {M}onte {C}arlo statistical
  inference.
\newblock {\em Foundations and Trends in Machine Learning}, 6(1):1--143, 2013.

\bibitem{LindstenSO:2011}
F.~Lindsten, T.~B. Sch\"on, and J.~Olsson.
\newblock An explicit variance reduction expression for the
  {R}ao-{B}lackwellised particle filter.
\newblock In {\em Proceedings of the 18th {IFAC} World Congress}, Milan, Italy,
  August 2011.

\bibitem{Liu:2001}
J.~S. Liu.
\newblock {\em {M}onte {C}arlo Strategies in Scientific Computing}.
\newblock Springer, 2001.

\bibitem{Logothetis99}
A.~Logothetis and V.~Krishnamurthy.
\newblock Expectation maximization algorithms for map estimation of jump
  {M}arkov linear systems.
\newblock {\em Signal Processing, IEEE Transactions on}, 47(8):2139 --2156, aug
  1999.

\bibitem{Mazor1998}
E.~Mazor, A.~Averbuch, Y.~Bar-Shalom, and J.~Dayan.
\newblock Interacting multiple model methods in target tracking: A survey.
\newblock {\em Aerospace and Electronic Systems, IEEE Transactions on},
  34(1):103--123, 1998.

\bibitem{mclachlan2007}
G.~McLachlan and T.~Krishnan.
\newblock {\em The {EM} algorithm and extensions}, volume 382.
\newblock John Wiley \& Sons, 2007.

\bibitem{Medbo2009}
J.~Medbo, I.~Siomina, A.~Kangas, and J.~Furuskog.
\newblock Propagation channel impact on {LTE} positioning accuracy - {A} study
  based on real measurements of observed time difference of arrival.
\newblock In {\em Proc. of IEEE International Symposium on Personal, Indoor and
  Mobile Radio Communications}, pages 2213 -- 2217, Tokyo, Japan, Sep. 2009.

\bibitem{Mihaylova2007}
L.~Mihaylova, D.~Angelova, S.~Honary, D.R. Bull, C.~N. Canagarajah, and
  B.~Ristic.
\newblock Mobility tracking in cellular networks using particle filtering.
\newblock {\em IEEE Transactions on Wireless Communication}, 6:3589--3599, Oct
  2007.

\bibitem{Mongillo2008}
G.~Mongillo and S.~Denève.
\newblock Online learning with hidden {Markov} models.
\newblock {\em Neural Computation}, 20(7):1706--1716, 2008.

\bibitem{munir1995adaptive}
A.~Munir and D.P. Atherton.
\newblock Adaptive interacting multiple model algorithm for tracking a
  manoeuvring target.
\newblock {\em IEE Proceedings-Radar, Sonar and Navigation}, 142(1):11--17,
  1995.

\bibitem{Nicoli2008}
M.~Nicoli, C.~Morelli, and V.~Rampa.
\newblock A jump {M}arkov particle filter for localization of moving terminals
  in multipath indoor scenarios.
\newblock {\em Signal Processing, IEEE Transactions on}, 56(8):3801--3809,
  2008.

\bibitem{Orguner2006online}
U.~Orguner and M.~Demirekler.
\newblock An online sequential algorithm for the estimation of transition
  probabilities for jump {M}arkov linear systems.
\newblock {\em Automatica}, 42(10):1735--1744, 2006.

\bibitem{Orguner2008}
U.~Orguner and M.~Demirekler.
\newblock {M}aximum {L}ikelihood estimation of transition probabilities of jump
  {M}arkov linear systems.
\newblock {\em {IEEE} Transactions on Signal Processing}, 56(10):5093 --5108,
  October. 2008.

\bibitem{ozkan2012}
E.~Ozkan, C.~Fritsche, and F.~Gustafsson.
\newblock Online {EM} algorithm for joint state and mixture measurement noise
  estimation.
\newblock In {\em Information Fusion (FUSION), 2012 15th International
  Conference on}, pages 1935--1940. IEEE, 2012.

\bibitem{PittS:1999}
M.~K. Pitt and N.~Shephard.
\newblock Filtering via simulation: Auxiliary particle filters.
\newblock {\em Journal of the American Statistical Association},
  94(446):590--599, 1999.

\bibitem{SchonGN:2005}
T.~Sch\"{o}n, F.~Gustafsson, and P.-J. Nordlund.
\newblock Marginalized particle filters for mixed linear/nonlinear state-space
  models.
\newblock {\em {IEEE} Transactions on Signal Processing}, 53(7):2279--2289,
  July 2005.

\bibitem{shumway1991dynamic}
RH~Shumway and DS~Stoffer.
\newblock Dynamic linear models with switching.
\newblock {\em Journal of the American Statistical Association},
  86(415):763--769, 1991.

\bibitem{Tugnait1982}
J.~Tugnait.
\newblock Adaptive estimation and identification for discrete systems with
  {M}arkov jump parameters.
\newblock {\em Automatic Control, IEEE Transactions on}, 27(5):1054--1065,
  1982.

\bibitem{vo2006}
Ba-Ngu Vo, A.~Pasha, and H.~D. Tuan.
\newblock A gaussian mixture {PHD} filter for nonlinear jump {M}arkov models.
\newblock In {\em Decision and Control, 2006 45th IEEE Conference on}, pages
  3162--3167. IEEE, 2006.

\bibitem{WhiteleyJ:2013}
N.~Whiteley and A.~Johansen.
\newblock Recent developments in auxiliary particle filtering.
\newblock In {\em Inference and Learning in Dynamic Models (to appear)}.
  Cambridge University Press, 2013.

\bibitem{Wu:1983}
C.~F.~J. Wu.
\newblock On the convergence properties of the {EM} algorithm.
\newblock {\em The Annals of Statistics}, 11(1):95--103, 1983.

\end{thebibliography}

% that's all folks
\end{document}